\documentclass[intlimits,twoside,a4paper]{article}
\usepackage[cp1251]{inputenc}

\usepackage[eqsecnum]{cmpj3}

\DeclareMathOperator{\sign}{sign}

\usepackage{bm}

\issue{2020}{23}{1}{13704}
\doinumber{10.5488/CMP.23.13704}

\title[The energy spectrum and the electrical conductivity of graphene with substitution impurity]%
{The energy spectrum and the electrical conductivity of graphene with substitution impurity}%

\author[S.P. Repetsky \textsl{et al.}]{S.P. Repetsky\refaddr{label1}, I.G. Vyshyvana\refaddr{label1}, S.P. Kruchinin\refaddr{label2},
R.M. Melnyk\refaddr{label3}, A.P. Polishchuk\refaddr{label4}}
\addresses{
\addr{label1} Institute of High Technologies, Taras Shevchenko National University of Kyiv,\\ 4-g Academician Glushkov Ave., 03022 Kyiv, Ukraine 
\addr{label2} Bogolyubov Institute for Theoretical Physics of the National Academy of Sciences of Ukraine,\\ 14-b Metrolohichna St., 03143 Kyiv, Ukraine 
\addr{label3} National University of Kyiv-Mohyla Academy, 2 Skovorody St., 04070 Kyiv, Ukraine 
\addr{label4} Aerospace Faculty, National Aviation University, 1 Kosmonavta Komarova Ave., 03058 Kyiv, Ukraine 
}

\date{Received August 28, 2019, in final form November 25, 2019}

\begin{document}

\maketitle

\begin{abstract}
The effect of substitution atoms on the energy spectrum and the electrical conductivity of graphene was investigated in a Lifshitz one-electron tight-binding model.
It is established that the ordering of impurity atoms results in a gap in the energy spectrum of electrons whose width depends on the order parameter and on the magnitude of the scattering potential. 
It is shown that if the order parameter is close to its maximum value, there are peaks associated with localized impurity states on the energy curve of the electron states density at the edges of the energy gap. 
At the electron concentration at which the Fermi level enters the gap region, the electrical conductivity is zero, and the metal-dielectric transition occurs. 
If the Fermi level falls in the region of the energy band, the electron relaxation time and electrical conductivity tend to infinity when the order parameter reaches its maximum value. 
The analytical calculations of the electron density and of the electrical conductivity of graphene, made in the limiting case of weak scattering, are compared with the results of the numerical calculations for different scattering potentials.

\end{abstract}

\section{Introduction}

Lately, the possibility of a targeted modification of graphene with the help of purposely-introduced impurities, created defects, and atoms or chemical functional groups deposited on a surface has attracted a particular attention. In this case, broad potentialities for a change in the physical properties of graphene open up due to the controlled introduction of impurities using the method of ionic implantation. Thus, graphene becomes a basis generating a new class of functional materials. Such materials sometimes find unexpected applications in nanoelectromechanical systems, systems of accumulation of hydrogen, etc. Surely, the main hopes are laid on graphene in connection with the expectation that in the nearest future it has all capabilities of becoming a successor of silicon in electronic devices, which permits to significantly increase the level of miniaturization and working frequencies thereof. The quasirelativistic spectrum of charge carriers underlines the uniqueness of graphene, but hampers, at the same time, the use of graphene in field transistors due to the absence of a gap in its spectrum. It is known that the impurities can induce the appearance of a gap in the energy spectrum. The width of the gap depends on the type of impurities and their concentration.

Most studies of the energy spectrum of graphene are based on the density functional theory. The most significant achievements are related to the self-consistent meta-gradient approximation and to the method of projection of adjoint waves \cite{Sun1} realized in the VASP and GAUSSIAN softwares \cite{Sun1}. The numerical calculations executed by this method showed the opening of a gap in the energy spectrum of graphene caused by the presence of an impurity. However,  clarification of the nature of this effect requires, in addition to the mentioned numerical calculations, the analytical studies of the influence of impurities on the energy spectrum and on the properties of graphene.

The electronic structures of the isolated monolayer of graphene, two- and three-layer graphenes and graphene grown on ultrathin layers of hexagonal boron nitride (h-BN) were calculated in \cite{Yelgel2} within the framework of the density functional theory with the use of the method of pseudopotential. It was shown that the energy gap $57$~meV in width appears in the graphene grown on a monolayer of h-BN.

Graphenes with impurities of aluminium, silicon, phosphorus, and sulphur were studied within the analogous method in work \cite{Denis3}, where it was shown, in particular, that graphene with a 3\% impurity of phosphorus has a gap $0.67$~eV in width. In work \cite{Xiaohui4} with the use of the QUANTUM-ESPRESSO software, the possibilities of the opening of a gap in the energy spectrum of graphene at the introduction of the impurities of boron and nitrogen (the gap width is $0.49$~eV), as well as the impurities of atoms of boron and atoms of lithium adsorbed on the surface (the gap width is $0.166$~eV), were demonstrated. 

It is obvious that it is insufficient to restrict ourselves by numerical calculations in order to understand the nature of the influence of impurities on the energy spectrum and properties of graphene. They should be also described within a simple but appropriate model that presents  exact analytical solutions.

In the Lifshitz tight-binding one-electron model, the theory of reconstruction of the spectrum of graphene caused by an increase in the concentration of point impurities was developed in works \cite{Skrypnyk5, Skrypnyk6, Pershoguba7, SkrypnykYu}. Moreover, the possibility of the metal-dielectric transition in such a system was predicted. The results of the analytical consideration of a reconstruction of the spectrum were confirmed with the help of a numerical experiment. This made it possible to verify the existence of a quasigap filled with localized states and showed its dominant role in the localization of the scattering by pairs and triples of impurity centers.

In \cite{yet1, yet2}, the splitting in the energy spectrum of graphene with a zigzag boundary was studied. This spectrum describes the electron waves that propagate along the boundary and decay with an increasing distance from it. It is shown that the electronic spectra of graphene with isolated vacancies exhibit a similar behaviour. The split electronic energy spectrum is accompanied by the formation of a sharp resonance state on the local density curve. It was shown that a similar resonance also arises in the phonon spectrum near the intersection point of the acoustic and optical branches of elastic waves polarized perpendicular to the plane of the graphene monolayer. In the frequency range under consideration, these phonons practically do not interact with differently polarized phonons, and also have high group velocities and make a dominant contribution to the electron-phonon interaction. The presented results demonstrate the possibility of increasing the critical temperature of the superconducting transition in graphene by a controlled creation of defects, such as a vacancy or a zigzag border.

The numerical calculations within the Kubo-Greenwood quantum-mechanical formalism in the Lifshitz tight-binding one-electron model were performed in \cite{Radchenko8, Radchenko9, Radchenko10, Radchenko11, Radchenko12, Radchenko13} to study the influence of impurity atoms or atoms adsorbed on the surface on the electronic structure and electrical conductance of graphene. In those works, the method of reducing the Hamiltonian to the three-diagonal form was developed to study the influence of completely ordered impurity atoms on the energy spectrum and electrical conductance of graphene in the ballistic and diffusive modes of conductance. In work \cite{Radchenko10}, it was found that the gap $ 0.45$~eV in width appears in the energy spectrum of electrons of graphene deposited on a potassium substrate. It was assumed in the paper that the appearance of this gap is associated with a change in the symmetry of the crystal. This assumption was corroborated in work \cite{Los14}, where the influence of the atomic ordering on the energy spectrum and electrical conductance of an alloy was analytically studied in the Lifshitz tight-binding one-electron model. It was also established \cite{Los14} that, for a long-range ordering of the alloy, the gap arises in the energy spectrum of electrons. The gap width is equal to the difference of the scattering potentials of components of the alloy. It was also found that metal-dielectric transition appears in the alloy provided the Fermi level falls in the domain of the gap at a long-range atomic ordering. 

It is worth to note that the velocity of an electron at the Fermi level can decrease as Fermi level falls in the domain of the gap. Moreover, when the gap appears in the energy spectrum of graphene for the case where the Fermi level falls in the domain of the gap, the velocity of an electron at the Fermi level can decrease. This leads to a decrease in the mobility and electrical conductance of electrons, which can worsen the functional characteristics of graphene as a material for field transistors.

Within the Lifshitz tight-binding one-electron model, the influence of the ordering of impurities on the energy spectrum and electrical conductance of graphene was considered in work \cite{Repetsky15}. It was established that the ordering of substitutional atoms on the nodes of the crystal lattice causes the appearance of a gap $\eta  |\delta | $ in width in the energy spectrum of graphene centered at the point $y  \delta$, where $\eta $ is the ordering parameter, $\delta $ is the difference of the scattering potentials of impurity atoms and carbon, and $y$ is the impurity concentration. If the Fermi level falls in the domain of the gap, then the electrical conductance $ \sigma _{\alpha \alpha } \to \infty $ at the ordering of graphene, i.e., the metal-dielectric transition arises. If the Fermi level is located outside the gap, then the electrical conductance increases with the order parameter $\eta $ according to the rule $\sigma _{\alpha \alpha } \sim ( y^{2} - \frac{1}{4} \eta ^{2} )^{-1} $. As the ordering of impurity atoms $ \eta \to 1 $ at the concentration $y=1/2$, the electrical conductance of graphene $\sigma _{\alpha \alpha } \to \infty $, i.e., graphene transits to the state of ideal conductance. We note that the conclusions in work \cite{Repetsky15} were based on the results of analytical studies of the energy spectrum and electrical conductance of graphene performed in the approximation of coherent potential. However, the convergence region of the decomposition cluster used in \cite{Repetsky15} for the Green function, and the applicability of the coherence potential approximation have not been analyzed. Both themes are studied in the present work. The features of the energy spectrum of electrons in the region of the gap arising upon the ordering of impurity atoms are investigated herein.

\section{Theoretical model}

The Hamiltonian in a one-electron Lifshitz strong bond model describing single-electron states of graphene with substitutional impurity can be represented as \cite{Repetsky15}
		\begin{equation} \label{Eqq1} 
			H	= \sum _{ni}{\left| ni \right\rangle} v_{ni} {\left\langle ni \right|}
				+ \sum _{ni,n'i'\ne ni}{\left| ni \right\rangle} h_{ni,n'i'} {\left\langle n'i' \right|}  ,     
		\end{equation} 
where $h_{ni,n'i'} $ is a non-diagonal matrix element of the Hamiltonian (jump integral) in the Vane representation, which in the assumed approximation of the diagonal disorder is independent of the random distribution of atoms. The diagonal matrix element $v _{ni} $ is $v^\text{A} $ or $v^\text{B} $ depending on whether atom A or B is at the node $ni$, $n$ is the number of elementary cell, $i$ is the number of the sublattice node in the unit cell.

Add and subtract in expression \eqref{Eqq1} the translational invariant operator $\sum_{ni}{\left| ni \right\rangle} \sigma _{i} {\left\langle ni \right|} $, where $\sigma_{i} $ is the diagonal matrix element of the Hamiltonian of some effective ordered medium (coherent potential), which depends on the sublattice number. As a result, the graphene Hamiltonian can be represented as follows:
		\begin{align} \label{Eqq2}
			H					& =  \tilde{H}+\tilde{V} , \nonumber \\
			\tilde{H} & =  \sum _{ni}{\left| ni \right\rangle} \sigma _{i} {\left\langle ni \right|}  
										+\sum _{ni,n'i'\ne ni}{\left| ni \right\rangle} 
										h_{ni,n'i'}{\left\langle n'i'\right|} , \nonumber \\
			\tilde{V} & =  \sum _{ni}\tilde{v}_{ni}\,  , \qquad 
											\tilde{v}_{ni} 	=  {\left| ni \right\rangle} (v_{ni} 
											- \sigma _{i} ){\left\langle ni \right|}. 
		\end{align}

The Green's belated function is an analytic function in the upper half-plane of the complex energy $z$ values. The function is determined by the expression
		\begin{equation} \label{Eqq3} 
		G(z)=(z-H)^{-1}.       
		\end{equation} 
Green's function satisfies the equation
		\begin{equation} \label{Eqq4} 
		G=\tilde{G}+\tilde{G}T\tilde{G},       
		\end{equation} 
Green's function $\tilde{G}$ of the effective medium corresponds to the Hamiltonian $\tilde{H}$ in \eqref{Eqq2}. The $T$ in the scattering  matrix can be represented as an infinite series \cite{Los14}
		\begin{equation} \label{Eqq5} 
		T = \sum _{(n_{1} i_{1} )}t^{n_{1} i_{1} }  
			+	\sum _{(n_{1} i_{1} )\ne (n_{2} i_{2} )}^{}T^{(2){\rm \; }n_{1} i_{1} ,n_{2} i_{2} } + \dots \,.
		\end{equation} 
Here,
		\begin{equation} \label{Eqq6} 
			T^{(2)  n_{1} i_{1} ,  n_{2} i_{2} } 
				=	\left[ I - t^{n_{1}i_{1}}  \tilde{G}  t^{n_{2}i_{2}}  \tilde{G} \right]^{-1}
					 t^{n_{1}i_{1} } 	\tilde{G}  t^{n_{2}i_{2}}  \left[ I + \tilde{G} 
					 t^{n_{1}i_{1}} \right] 
		\end{equation} 
and the scattering operator on one node
		\begin{equation} \label{Eqq7} 
			t^{n_{1}i_{1}} = \left[ I - \tilde{v}_{i n} \tilde{G} \right]^{-1}  \tilde{v}_{i n}\,,       
		\end{equation} 
$I$ is the identity matrix. The members of series \eqref{Eqq5} describe the processes of multiple scattering of electrons in clusters of one, two, three, etc. scattering centers. 

As was shown in work \cite{Los14}, the contributions of electrons scattering processes on clusters to the density of states and to the electrical conductance decrease as the number of atoms in a cluster increases. These contributions are guided by some small parameter $p_{i} (\varepsilon )$. The parameter $p_{i} (\varepsilon )$ is small in a wide region of the changes of crystal characteristics, except for narrow energy intervals on the edges of the spectrum and on the edges of the energy gap. The expression for the specified $p_{i} (\varepsilon )$ parameter is shown below.

Neglecting the contribution of scattering processes on clusters of three or more atoms that are small by the specified parameter $p_{i} (\varepsilon )$, the density of one-electron states of graphene can be represented as \cite{Repetsky15}
		\begin{align} \label{Eqq8}
			g \left( \varepsilon \right)			& = 		
					\frac{1}{\nu} \sum _{i,\lambda } P_{}^{\lambda 0i} 
					g^{\lambda 0i}  \left(\varepsilon \right), \nonumber \\
			g^{\lambda 0i} (\varepsilon ) 	& = 
					-	\frac{2}{\piup } {\Im  } \Bigl\{ \Bigr.  \tilde{G}
					+	\tilde{G}  t_{}^{\lambda 0i}  \tilde{G}  
					+ \sum _{\left(lj\right) \ne \left(0i\right),  \lambda'}
						P^{\lambda'  lj  / \lambda 0  i  } \tilde{G} \left[t^{\lambda' l j}  
					+ T^{(2)\lambda 0i,  \lambda 'lj} \right]
					 \tilde{G}^{ } \Bigl. \Bigr\}_{0i,0i},
		\end{align}
$\nu =2$ is the number of graphene sublattices.

Using the Cuban-Greenwood formula and neglecting the contribution of scattering processes on clusters of three or more atoms, we present the static electrical conductivity of graphene ($T=0$) \cite{Repetsky15} as follows:
\begin{align} \label{Eqq9} 
	\sigma_{\alpha \beta}	& =  
						- \frac{e^{2} \hbar}{2 \piup \Omega_{1}} 	\sum_{s,s'=+,-}(2 \delta_{ss'} - 1) \sum_{i} 
	 \Bigl( \Bigr. 
							[v_{\beta} \tilde{K}(\varepsilon^{s}, v_{\alpha},\varepsilon_{}^{s'})]  \nonumber \\ &   
	+ \sum_{\lambda} P^{\lambda 0 i} 
							\tilde{K}(\varepsilon^{s'}, v_{\beta}, \varepsilon ^{s}) t^{\lambda 0 i}(\varepsilon^{s})
							\tilde{K}(\varepsilon^{s}, v_{\alpha}, \varepsilon^{s'}) t^{\lambda 0 i}(\varepsilon^{s'}) \nonumber \\	&  
	+ \sum_{\lambda} P^{\lambda 0 i}  \sum_{lj \ne 0i,  \lambda'}  P^{\lambda' l j / \lambda 0 i}   
	 \Bigl\{ \Bigr. 	
							[\tilde{K}(\varepsilon^{s'}, v_{\beta}, \varepsilon^{s}) v_{\alpha} \tilde{G}(\varepsilon^{s'})]
							 	T^{(2) \lambda 0 i, \lambda' l j}(\varepsilon^{s'})  \nonumber \\ &  
						+	 [\tilde{K}(\varepsilon^{s} ,v_{\alpha}, \varepsilon^{s'}) v_{\beta} \tilde{G}(\varepsilon^{s})]
							  T^{(2) \lambda 0 i, \lambda' l j}(\varepsilon^{s})  
						+  \tilde{K}(\varepsilon^{s'}, v_{\beta}, \varepsilon^{s})
	 \Bigl[ \Bigr.  
							t^{\lambda 'lj} (\varepsilon^{s} )	
							\tilde{K}(\varepsilon^{s},v_{\alpha},\varepsilon^{s'})
							t_{}^{\lambda 0i}(\varepsilon^{s'})  \nonumber \\ & 
						+  [t_{0 i}^{\lambda}(\varepsilon^{s}) + t_{l j}^{\lambda'}(\varepsilon^{s})]
							\tilde{K}(\varepsilon^{s},v_{\alpha},\varepsilon^{s'}) 
							T^{(2) \lambda 0 i, \lambda' l j}(\varepsilon^{s'})  
						+	 T^{(2) \lambda' l j, \lambda 0 i}(\varepsilon^{s})
							\tilde{K}(\varepsilon^{s}, v_{\alpha}, \varepsilon^{s'})
							t^{\lambda 0 i}(\varepsilon^{s'})  \nonumber \\ &  
						+  T^{(2) \lambda' l j, \lambda 0 i}(\varepsilon^{s})
							\tilde{K}(\varepsilon^{s}, v_{\alpha}, \varepsilon^{s'})
							T^{(2) \lambda 0 i, \lambda' l j}(\varepsilon^{s'})  \nonumber \\ & 
						+  T^{(2) \lambda' l j, \lambda 0 i}(\varepsilon^{s})
							\tilde{K}(\varepsilon^{s}, v_{\alpha}, \varepsilon^{s'})
							T^{(2) \lambda' l j, \lambda 0 i}(\varepsilon^{s'})
	 \Bigl. \Bigr]  \Bigl. \Bigr\}  \Bigl. \Bigr)_{0i,0i}  \Bigl. \Bigr|_{\varepsilon=\mu}. 
		\end{align}
 $\tilde{K}(\varepsilon_{}^{s}, v_{\alpha}, \varepsilon_{}^{s'}) 
 = \tilde{G}(\varepsilon_{}^{s}) v_{\alpha} \tilde{G}(\varepsilon^{s'})$, 
 	$\tilde{G}(\varepsilon^{+}) = \tilde{G}_\text{r}(\varepsilon)$, 
  	$\tilde{G}(\varepsilon_{1}^{-}) = \tilde{G}_\text{a}(\varepsilon) 
		= \tilde{G}_\text{r}^{*}(\varepsilon)$, 
 $\tilde{G}_\text{r}(\varepsilon)$, $\tilde{G}_\text{a}(\varepsilon)$ are retarded and advanced Green functions. 
 $\Omega_{1} = 2 \Omega_{0}$ is the graphene unit cell volume, $\Omega_{0}$ is the one atom volume.
		\begin{equation} \label{Eqq10} 
			\tilde{G}_{n j n' j'}(\varepsilon) 
			= \frac{1}{N} \sum _{\mathbf{k}}\tilde{G}_{j j'}(\mathbf{k}, \varepsilon) \exp [\ri \mathbf{k}(r_{n' j'} - r_{n j})] ,
		\end{equation} 
$\tilde{G}_{j j'}(\mathbf{k}, \varepsilon)$ is Fourier transform of Green's function, $r_{n j}$ is position vector of node $n j$. The wave vector $\mathbf{k}$ changes within the Brillouin zone.

The operator of electron velocity $\alpha$-projection is given by the expression
		\begin{equation} \label{Eqq11} 
			\upsilon_{\alpha i i'}(\mathbf{k}) = \frac{1}{\hbar } \frac{\partial h_{i i'}(\mathbf{k})}{\partial k_{\alpha}}\,,       
		\end{equation} 
Fourier transform of jump integral $h_{jj'}(\mathbf{k})$ is calculated for the nearest atom neighbours
		\begin{equation} \label{Eqq12} 
		h_{j j'} (\mathbf{k}) = \gamma_{1} \sum_{n' \ne n} \exp[\ri \mathbf{k} (r_{n' j'} - r_{n j})] ,     
		\end{equation} 
$\gamma_{1} = (p p \pi)$ is the jump integral \cite{Slater17}, $r_{n j}$ is position vector of node $n j$.

The Fermi level $\mu$ is determined from the equation
		\begin{equation} \label{Eqq13} 
		\left\langle Z \right\rangle = \int_{-\infty}^{\mu} g(\varepsilon)  \rd\varepsilon ,       
		\end{equation} 
here, $\left\langle Z \right\rangle$ is the average number of electrons per atom whose energy values belong to the energy band.

In expressions \eqref{Eqq8}, \eqref{Eqq9}, $P^{\lambda 0 i}$ is the filling probability of node $0 i$ of the crystal lattice $i = 1, 2$ by atoms of the sort $\lambda = \text{A, B}$
		\begin{equation} \label{Eqq14} 
		P^{\text{B}01} = y_{1} = y + \frac{1}{2} \eta, \quad P^{\text{B}02} = y_{2} = y - \frac{1}{2} \eta, \quad P^{\text{A}01} = 1 - P^{\text{B}01}. 
		\end{equation} 
$y$ is concentration of impurity atoms, $\eta$ is far order parameter.

In expressions \eqref{Eqq8}, \eqref{Eqq9} $P^{\lambda' l j / \lambda 0 i}$ 
is the probability of filling the node $l j$ by atom sort $\lambda '$ 
provided that the atom of the sort $\lambda$ fills the node $0i$. 
$P^{\lambda' l j / \lambda 0 i}$ is the parameter of paired interatomic correlations 
in atoms of crystalline lattice nodes filled with atoms.

The probabilities are determined by the interatomic pair correlations $\varepsilon_{l j, 0 i}^\text{BB}$ via
\cite{Repetsky20, Repetsky21}
		\begin{equation} \label{Eqq15} 
		P_{l j, 0 i}^{\lambda' / \lambda} 
		= P_{l j}^{\lambda'} + \frac{\varepsilon_{l j, 0 i}^\text{BB}}{P_{0 i}^{\lambda}} 
		(\delta_{\lambda' \text{B}} - \delta_{\lambda' \text{A}})  (\delta_{\lambda \text{B}} - \delta_{\lambda \text{A}}),     
		\end{equation} 
where $\delta$ is the Kronecker delta-function. Note that the interatomic pair correlations also satisfy 
		\begin{equation} \label{Eqq16} 
		\varepsilon_{l j, 0 i}^\text{BB}=\langle(c_{l j}^\text{B} - c_{j}^\text{B})(c_{0i}^\text{B} - c_{i}^\text{B})\rangle.       
		\end{equation} 
Here, $c_{lj}^\text{B}$ is bit randomized number which takes the value of one 
if the sort atom B is in the node or zero otherwise, 
$A_{j}^\text{B} = \langle A_{0 j}^\text{B} \rangle = P^{\text{B} 0 j}$. 
Brackets mean the averaging over the distribution of impurity atoms at the nodes of the crystalline lattice.

Coherent potential is determined by the condition $\langle t^{n_{1} i_{1} } \rangle =0$, hence the equation for the coherent potential \cite{Repetsky15} 
\begin{equation} \label{Eqq17} 
\sigma _{i} =\left\langle \upsilon _{i} \right\rangle -(\upsilon _\text{A} -\sigma _{i} ) \tilde{G}_{0i,0i} (\varepsilon ) (\upsilon _\text{B} -\sigma _{i} ); \quad  \left\langle \upsilon _{i} \right\rangle =(1-y_{i} ) \upsilon _\text{A}  +y_{i}  \upsilon _\text{B} . 
\end{equation} 

Putting $\upsilon _\text{A} =0$ in expression \eqref{Eqq17}, we obtain
\begin{equation} \label{Eqq18} 
\left\langle \upsilon _{i} \right\rangle =y_{i}  \delta ,         
\end{equation} 
here, 
\begin{equation} \label{Eqq19} 
\delta =\upsilon _\text{B} -\upsilon _\text{A} \,,         
\end{equation} 
the difference of scattering potentials of graphene components.

For analytical description of the energy spectrum and the electrical conductivity of graphene, we consider only the first constituents in expressions (\ref{Eqq8}), (\ref{Eqq9}), which make a major contribution to the density of states and to the electrical conductivity. Thus,
		\begin{align} 
		\label{Eqq20}
			g(\varepsilon )							& = 		
					-\frac{2}{\piup \nu } \Im\sum _{i}\tilde{G}_{0i,0i} (\varepsilon )
					=-\frac{2}{\piup \nu N} \Im\sum _{i,\mathbf{k}}\tilde{G}_{ii} (\mathbf{k},\varepsilon ), \\ 
		\label{Eqq21}
			\sigma _{\alpha \alpha }  	& = 
					-\frac{e^{2} \hbar }{2\piup V_{1} } \sum _{i}\left\{\upsilon _{\alpha } 
					[\tilde{G}(\varepsilon )-\tilde{G}^{*} (\varepsilon )]  
					\upsilon _{\alpha } [\tilde{G}(\varepsilon ) 
					-\tilde{G}^{*} (\varepsilon )]\right\} _{0i,0i}  \nonumber \\
																	& = 				
					-\frac{e^{2} \hbar }{2\piup V_{1} N} \sum _{i,\mathbf{k}}\left\{\upsilon _{\alpha }(\mathbf{k})
					[\tilde{G}(k,\varepsilon ) - \tilde{G}^{*} (\mathbf{k},\varepsilon )]  
					\upsilon _{\alpha } (\mathbf{k})[\tilde{G}(\mathbf{k},\varepsilon )
					-\tilde{G}^{*} (\mathbf{k},\varepsilon )]\right\} _{0i,0i}.
		\end{align}
The wave vector in formulae \eqref{Eqq20}, \eqref{Eqq21} varies within the Brillouin zone. Fourier transform of Green's function 
\begin{align} \label{Eqq22} 
		&\displaystyle	{\tilde{G}_{11} (\mathbf{k},\varepsilon )
		=\frac{\varepsilon -\sigma _{2} }{D(\mathbf{k},\varepsilon )} \,,    \quad 
		\tilde{G}_{12} (\mathbf{k},\varepsilon )=\frac{h_{21} (\mathbf{k})}{D(\mathbf{k},\varepsilon )} \,,} \nonumber\\ 
				&\displaystyle	{\tilde{G}_{21} (\mathbf{k},\varepsilon )
				=\frac{h_{12} (\mathbf{k})}{D(\mathbf{k},\varepsilon )}\, ,   \quad   		
				\tilde{G}_{22} (\mathbf{k},\varepsilon )
				=\frac{\varepsilon -\sigma _{1} }{\varepsilon -\sigma _{2} } \tilde{G}_{11} (\mathbf{k},\varepsilon ),} \nonumber\\ 
						&\displaystyle	{D(\mathbf{k},\varepsilon )
						=(\varepsilon -\sigma _{1} )(\varepsilon -\sigma _{2} )-h_{12} 			(\mathbf{k})h_{21} (\mathbf{k}).} 
\end{align} 
Fourier transform of the jump integral $h_{ii'} (\mathbf{k})$ is calculated for the nearest atom neighbours.

In this model, the value of the wave vector lying in the regions around the Dirac points is mainly due to the energy spectrum of electrons in the middle of the zone. The Brillouin zone has two areas of this kind. For these areas
		\begin{equation} \label{Eqq23} 
		h_{12} (\mathbf{k})=h_{21} (\mathbf{k})=\hbar \upsilon _\text{F} k, 
		\end{equation} 
$ \upsilon _\text{F} =\frac{3|\gamma _{1} |a_{0} }{2\hbar } $ is the electron velocity at the Fermi level, $\gamma _{1} =(pp\piup )$ is the jump integral \cite{Slater17}, $a_{0} $ is the distance between the nearest neighbours.

Substituting \eqref{Eqq22}, \eqref{Eqq23} in \eqref{Eqq20} and replacing the sum over the wave vector by an integral \cite{Repetsky15}
\begin{align} \label{Eqq24} 
&\displaystyle 
			{\tilde{G}_{01,01} (\varepsilon )
			=-\frac{S_{1} (\varepsilon -\sigma _{2} )}{\piup \hbar ^{2} \upsilon _\text{F}^{2} } 
			\ln \sqrt{1-\frac{w^{2} }{(\varepsilon -\sigma _{1} )(\varepsilon -\sigma _{2} )} } \,,} \nonumber\\ 
&\displaystyle  
		{\tilde{G}_{02,02} (\varepsilon )
		=-\frac{S_{1} (\varepsilon -\sigma _{1} )}{\piup \hbar ^{2} \upsilon _\text{F}^{2} } 
		\ln \sqrt{1-\frac{w^{2} }{(\varepsilon -\sigma _{1} )(\varepsilon -\sigma _{2} )} } \,,} 
\end{align}
$w=3|\gamma _{1} |$ is is the half-width of the energy band of pure graphene, $ S_{1} = \frac{3\sqrt{3} a_{0}^{2}}{2}$ is the area of the unit cell of graphene.

Consider the effect of the ordering atoms on the energy spectrum of graphene electrons with an admixture of substitution in the limiting case of weak scattering $\left|\delta /w\right|\ll 1$. In this case, the solution of the system of equations \eqref{Eqq17}, \eqref{Eqq24} is \cite{Repetsky15}
\begin{align} \label{Eqq25}
			&\displaystyle  {\tilde{G}_{01,01} (\varepsilon )
			=-\frac{S_{1} (\varepsilon -\sigma '_{2} )}{\piup \hbar ^{2} \upsilon _\text{F}^{2} } 
			\ln \sqrt{1-\frac{w^{2} }{(\varepsilon -\sigma '_{1} )(\varepsilon -\sigma '_{2} )} } \,,} \nonumber\\ 
					&\displaystyle  {\tilde{G}_{02,02} (\varepsilon )
					=-\frac{S_{1} (\varepsilon -\sigma '_{1} )}{\piup \hbar ^{2} \upsilon _\text{F}^{2} } 
					\ln \sqrt{1-\frac{w^{2} }{(\varepsilon -\sigma '_{1} )(\varepsilon -\sigma '_{2} )} } \,,} \nonumber\\ 
							&\displaystyle  {\sigma '_{1} = y_{1} \delta -y_{1} (1-y_{1} )\delta ^{2}  
							\frac{S_{1} (\varepsilon -y_{2} \delta )}{\piup \hbar ^{2} \upsilon _\text{F}^{2} } 
							\ln \sqrt{1-\frac{w^{2} }{(\varepsilon -y_{1} \delta )(\varepsilon -y_{2} \delta )} } \,,} \nonumber\\ 
									&\displaystyle  {\sigma '_{2} =y_{2} \delta -y_{2} (1-y_{2} )\delta ^{2}  
									\frac{S_{1} (\varepsilon -y_{1} \delta )}{\piup \hbar ^{2} \upsilon _\text{F}^{2} } 
									\ln \sqrt{1-\frac{w^{2} }{(\varepsilon -y_{1} \delta )(\varepsilon -y_{2} \delta )} } \,,} \nonumber\\ 
											&{\sign (\varepsilon -\sigma '_{1} )=-\sign (\varepsilon -\sigma '_{2} ); } 
\end{align}
and
\begin{align}  
		&\displaystyle  {\tilde{G}_{01,01} (\varepsilon )
		=-\frac{S_{1} (\varepsilon -\sigma '_{2} )}{\piup \hbar ^{2} \upsilon _\text{F}^{2} } 
		\ln \sqrt{\frac{w^{2} }{(\varepsilon -\sigma '_{1} )(\varepsilon -\sigma '_{2} )} -1} 
		-\ri\frac{S_{1} |\varepsilon -\sigma '_{2} |}{2\hbar ^{2} \upsilon _\text{F}^{2} } \,,} \nonumber\\ 
				&\displaystyle  {\tilde{G}_{02,02} (\varepsilon )
				=-\frac{S_{1} (\varepsilon -\sigma '_{1} )}{\piup \hbar ^{2} 
				\upsilon _\text{F}^{2} } \ln \sqrt{\frac{w^{2} }{(\varepsilon -\sigma '_{1} )(\varepsilon -\sigma '_{2} )} -1} 
				-\ri\frac{S_{1} |\varepsilon -\sigma '_{1} |}{2\hbar ^{2} \upsilon _\text{F}^{2} }\, ,} \nonumber \\ 
				&\displaystyle  {\sigma '_{1} =y_{1} \delta -y_{1} (1-y_{1} )\delta ^{2}  
						\frac{S_{1} (\varepsilon -y_{2} \delta )}{\piup \hbar ^{2} 
						\upsilon _\text{F}^{2} } \ln \sqrt{\frac{w^{2} }{|(\varepsilon -y_{1} \delta )(\varepsilon -y_{2} \delta )|} -1} ,} \nonumber  
\end{align} 
\begin{align}   \label{Eqq26}
				&\displaystyle  {\sigma ''_{1} =-y_{1} (1-y_{1} )\delta ^{2}  \frac{S_{1} 
								|\varepsilon -y_{2} \delta |}{2\hbar ^{2} \upsilon _\text{F}^{2} }\, ,} \nonumber \\
						&\displaystyle {\sigma '_{2} =y_{2} \delta -y_{2} (1-y_{2} )\delta ^{2}  
						\frac{S_{1} (\varepsilon -y_{1} \delta )}{\piup \hbar ^{2} \upsilon _\text{F}^{2} } 
						\ln \sqrt{\frac{w^{2} }{|(\varepsilon -y_{1} \delta )(\varepsilon -y_{2} \delta )|} -1} ,} \nonumber \\ 
								&\displaystyle {\sigma ''_{2} =-y_{2} (1-y_{2} )\delta ^{2}  
								\frac{S_{1} |\varepsilon -y_{1} \delta |}{2\hbar ^{2} \upsilon _\text{F}^{2} }\, ,} \nonumber\\
						&{\sign (\varepsilon -\sigma '_{1} )=\sign (\varepsilon -\sigma '_{2} ).} 
\end{align} 
In (\ref{Eqq25})--(\ref{Eqq26}) equations, $\sigma '_{i} $ and $\sigma ''_{i} $ are the real and imaginary parts of the coherent potentials $\sigma _{i}$,  $i=1,2$.

The analysis of formulae (\ref{Eqq25})--(\ref{Eqq26}) shows that if the
impurity atoms are ordered in the crystal lattice of graphene then in the
energy spectrum there is a gap of width $\eta | \delta |$ and a center of
gap is at $y\delta $. 
The energy values of $\varepsilon $ corresponding to the edges of the energy gap are determined from the equations $\varepsilon -\sigma '_{1} =0$, $\varepsilon -\sigma '_{2} =0$. It follows from equations \eqref{Eqq14} that the maximum value of the ordering parameter $\eta _{\max } =2y$,  $y<1/2$. At the complete ordering of the impurity atoms, the width of the gap is equal to $2y|\delta |$, proportional to the concentration of the impurity $y$ and the modulus of scattering potential of the graphene components  $\delta $. For $y=1/2$, the width of the gap takes the maximum value. For $\delta >0$ and $\delta <0$, the gap is located respectively to the right and to the left of the Dirac point on the energy scale. As can be seen from formulae (\ref{Eqq20}), (\ref{Eqq25}), the density of electron states $g(\varepsilon )=0$ in the approximation of the coherent potential for this region of energy values.

As it follows from expressions (\ref{Eqq20}), (\ref{Eqq26}), in the vicinity of the edge of the gap, the density of states tends to infinity. This is due to the presence of the second components in the expressions for coherent potentials $\sigma '_{1} $, $\sigma '_{2} $ \eqref{Eqq26}. The width of this energy region is \cite{Repetsky15}
\begin{equation} \label{Eqq27} 
\left|\frac{\Delta \varepsilon (\eta )}{w} \right|= \frac{w}{\eta |\delta |} \exp \left[-\frac{2y\piup w^{2} }{3\sqrt{3} \eta \delta ^{2} (1-y+\eta /2)(y-\eta /2)} \right]; \quad 0 < \eta \leqslant 2y. 
\end{equation} 
The peak width estimation \eqref{Eqq27} is made on condition that the density of states on the slope of the peak is twice its value at the point of the adjacent minimum.

Beyond the specified peak, the density of states increases linearly with an increasing distance to the gap edge \cite{Repetsky15}
\begin{equation} \label{Eqq28} 
g(\varepsilon )=\frac{S_{1} (\varepsilon -y\delta )}{\piup \hbar ^{2} \upsilon _\text{F}^{2} }\, ,   \quad    \left|\frac{\Delta \varepsilon (\eta )}{w} \right|<\left|\frac{\varepsilon -\sigma '_{i} }{w} \right|\leqslant \left|\frac{\delta }{w} \right|. 
\end{equation} 

If the Fermi level falls into the gap region, then the number of free charge carriers tends to zero. In this case, when the impurity is ordered, the electrical conductivity $\sigma _{\alpha \alpha } \to 0$, as it follows from formulae~(\ref{Eqq21}), (\ref{Eqq25}), i.e., a metal-dielectric transition occurs.

Let us explore the electrical conductivity of graphene when the Fermi level is outside the gap. Substituting \eqref{Eqq22}, \eqref{Eqq23} in \eqref{Eqq20} and replacing the sum over the wave vector by an integral \cite{Repetsky15}, we obtain
\begin{equation} \label{Eqq29} 
\sigma _{\alpha \alpha } =\frac{2e^{2} \hbar \upsilon _\text{F}^{2} }{\piup ^{2} a_{0}^{2} d\left(y^{2} -\frac{1}{4} \eta ^{2} \right)\delta ^{2} } .       
\end{equation} 
Where $d$ is the thickness of graphene. The $d$ factor in the denominator of the right-hand side of formula~\eqref{Eqq29} can be omitted, since the expression for the electrical resistance of graphene is reduced.

In \cite{Los14} it is shown that the contribution of electron scattering processes on clusters to the density of states and the electrical conductance is guided by some small parameter $p_{i} (\varepsilon )$, except for narrow energy intervals on the edges of the spectrum and on the edges of the energy gap. In \cite{Repetsky15}, an analytical study of the effect of impurity ordering on the energy spectrum and the electrical conductance of graphene in the approximation of coherent potential was executed. In order to estimate the corrections of this approximation caused by the contribution of electron scattering processes by clusters of two, three, etc. atoms, the following parameter was introduced in \cite{Ducastelle19}
\begin{align} \label{Eqq30}  &{p_{i} 
		\displaystyle		\left(\varepsilon \right)
		=\bigg|\left\langle [t^{0i} \left(\varepsilon \right)]^{2} \right\rangle \sum _{lj\ne 0i}
		\tilde{G}_{0i,lj} \left(\varepsilon \right)\tilde{G}_{lj,0i} \left(\varepsilon \right) \bigg|;} \nonumber\\ 
				&{\left\langle [t^{0i} \left(\varepsilon \right)]^{2} \right\rangle 
				=\left(1-y_{i} \right)[t^\text{A} {}^{0i} \left(\varepsilon \right)]^{2} +y_{i} [t^\text{B} {}^{0i} 
				\left(\varepsilon \right)]^{2} .}
\end{align} 
This parameter was analyzed in \cite{Los14}, where the following representation thereof was presented
\begin{align} \label{Eqq31} 
&{p_{i} \left(\varepsilon \right)=\displaystyle\left|\frac{Q_{i} (\varepsilon )}{1+Q_{i} (\varepsilon )}  \right|;} \nonumber\\ 
		&\displaystyle  {Q_{i} (\varepsilon )
		=-\frac{\left\langle [t^{0i} (\varepsilon )]^{2} \right\rangle }
			{1+\left\langle [t^{0i} (\varepsilon )]^{2} \right\rangle \left[\tilde{G}_{0i,0i} (\varepsilon )\right]^{2} } 					\left\{\frac{1}{1+\left\langle [t^{0i} (\varepsilon )]^{2} \right\rangle
					\left[\tilde{G}_{0i,0i} (\varepsilon )\right]^{2} }  \frac{\rd}{\rd\varepsilon }
					\tilde{G}_{0i,0i} (\varepsilon )+\left[\tilde{G}_{0i,0i} (\varepsilon )\right]^{2} \right\}.} 
\end{align} 

The parameter $p_{i}$ is small, with the exception of narrow intervals of energy values at the edges of the gap. As it follows from the formula \eqref{Eqq26}, the energy value tends to the energy of the edge of the gap $d\Im\tilde{G}_{0i,0i}(\varepsilon)/\varepsilon$ and the parameter $p_{i} \rightarrow \infty$ \eqref{Eqq31}.

The parameter $p_{i} (\varepsilon )$ takes on values $1/2\leqslant p_{i} (\varepsilon )\leqslant 1$ in a narrow range of energy values at the edge of the energy gap
\begin{equation} \label{Eqq32} 
\left|\frac{\Delta \varepsilon '(\eta )}{w} \right| = \frac{27}{\piup} \left(y^2 - \frac{1}{4} \eta^2\right)  
\left[(1 - y^2) - \frac{1}{4} \eta^2 \right]  \eta  \left(\frac{\delta}{w}\right)^5,
\end{equation} 
the expression is obtained by \eqref{Eqq7}, \eqref{Eqq26}, \eqref{Eqq31}.

Thus, the processes of scattering on clusters give a significant contribution to the density of states at
the energies of the electrons lying in the interval \eqref{Eqq19}. We note that formulae \eqref{Eqq28}, \eqref{Eqq29} for the density of states and
electrical conductance of graphene cannot be used, if the Fermi level falls in the interval of energies \eqref{Eqq32}
at the gap edges.   
 
The above expressions \eqref{Eqq28}, \eqref{Eqq29} were obtained for the case of a small value of the scattering potential $\left|\delta /w\right|\ll 1$. The influence of the ordering of impurity atoms on the energy spectrum and the electrical conductivity of graphene for an arbitrary value of the scattering potential is more complex.

\section{Results}

\begin{figure}[!t] 
\centerline{\footnotesize $\delta/w=-0.2$ \hspace{0.34\textwidth} $\delta/w=-0.6$}  \smallskip
\centerline{	\includegraphics[width=0.48\textwidth]{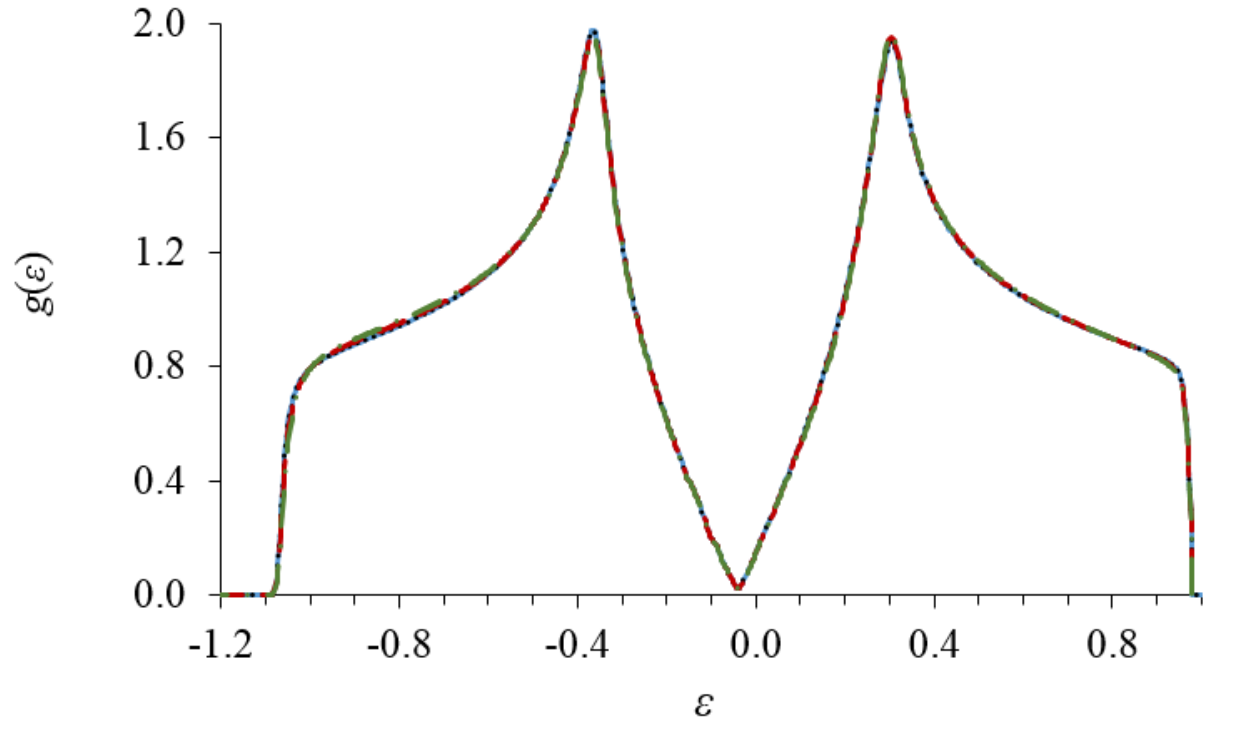} 	
							\includegraphics[width=0.48\textwidth]{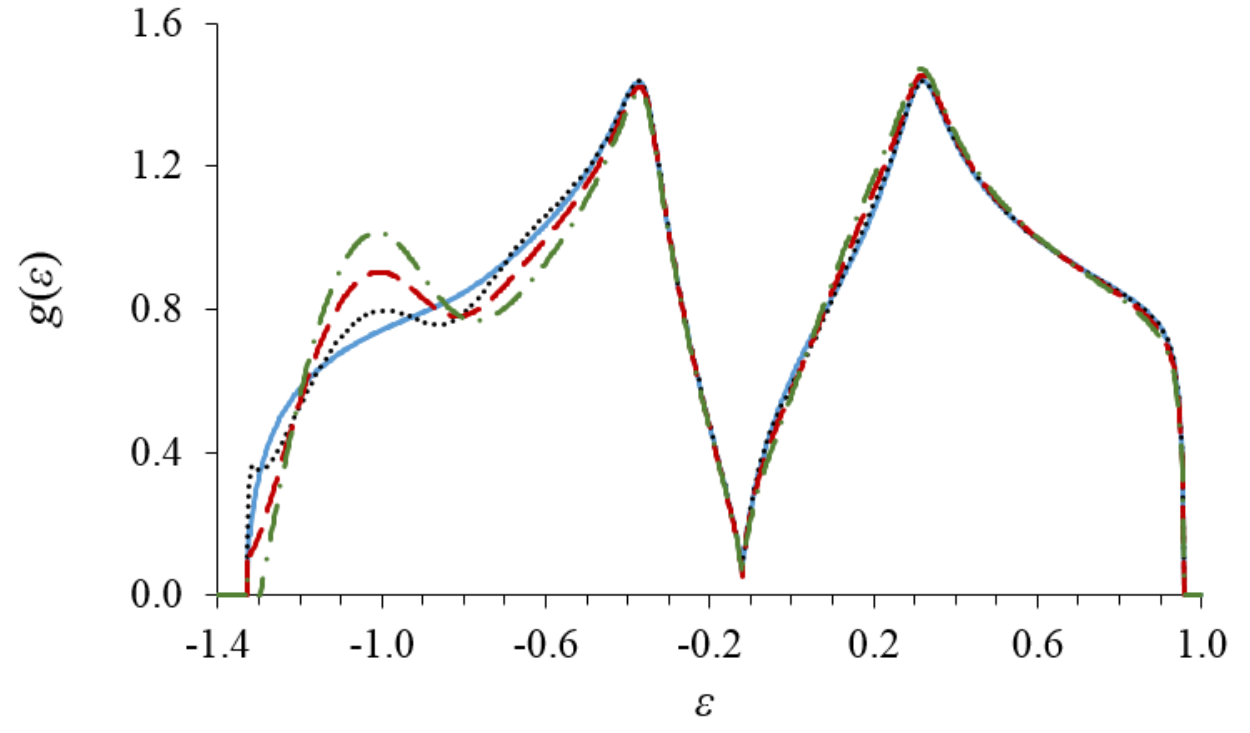}} \smallskip
\caption{\label{fig1} (Colour online)
The dependence of the density of electronic states $g(\varepsilon )$ on energy $\varepsilon $. 
The value of the concentration of substitutional impurity $y=0.2$, the scattering potential $\delta/w=-0.2$ on the left-hand side (lhs) 
and $\delta/w=-0.6$ on the right-hand side (rhs), the ordering parameter $\eta=0$, 
different values of the parameter of pair interatomic correlations $\varepsilon_{lj,0i}^\text{BB}=\varepsilon^\text{BB}$.
The density of states calculated in the approximation of the coherent potential (solid line) 
and taking account of the scattering processes on the pairs of atoms within the first coordination sphere:
completely disordered arrangement of impurity atoms $\varepsilon^\text{BB}=0$ (dotted line), 
with the interatomic pair correlations $\varepsilon^\text{BB}=-0.05$ (dashed line) 
and with $\varepsilon^\text{BB}=-0.1$ (dash-dotted line). } 
\end{figure}

Figure~\ref{fig1} shows the numerical calculation results of the density of state $g(\varepsilon)$ of graphene, performed according to \eqref{Eqq8} for the following values: the concentration of substitutional impurity $y=0.2$,  the scattering potential $\delta /w=-0.2$ and $\delta /w=-0.6$, the ordering parameter $\eta =0$ and different values of paired interatomic correlations in the first coordination sphere $\varepsilon _{lj  0i}^\text{BB} =\varepsilon _{}^\text{BB} $. Energy values are given in units of half-width of the energy band $w$. The electron density of states $g(\varepsilon )$ of graphene (figure~\ref{fig1}, solid curve) is calculated in the approximation of the coherent potential, where  only the first component of the sum in the formula \eqref{Eqq8} is taken into account. The dotted curve shows the behaviour of the density of states $g(\varepsilon )$ calculated with regard to the scattering processes of the pairs of atoms located within the first coordination sphere, for the case of completely disordered arrangement of impurity atoms on the graphene lattice, $\varepsilon _{}^\text{BB} =0$, $\eta =0$. The dotted curve and dash-dotted curve show the density of states $g(\varepsilon )$ calculated with regard to the scattering processes of the pairs of atoms if the paired interatomic correlation $\varepsilon^\text{BB}=-0.05$ and $\varepsilon^\text{BB}=-0.1$, the order parameter $\eta =0$. The curve describing the density of electron states in the approximation of the coherent potential and the curves taking account of  the scattering processes on the pairs of atoms  coincide in the case of small values of the scattering parameter (figure~\ref{fig1}, $\delta /w = -0.2$). In the case $\delta/w=-0.6$ (on the right-hand side figure~\ref{fig1}), there occurs a characteristic dip of the density of state, whose value increases with an increase of the correlation parameter $\varepsilon^\text{BB}$.

The values of the density of states $g(\varepsilon )$ were also calculated with respect to the scattering processes on the pairs of atoms, which are located within three coordination spheres and within ten coordination spheres. The results practically coincide with the results of calculations that take into account the scattering on the pairs within the first coordination sphere (figure~\ref{fig1}). Hence, we can conclude that the area of the influence of impurity electronic states of graphene in the specified model is limited by the first coordination sphere for the case of the concentration of substitutional impurity not more than $y=0.2$.

\begin{figure}[!t]
\centerline{\footnotesize $\delta/w=-0.2$ \hspace{0.34\textwidth} $\delta/w=-0.6$}  \smallskip
\centerline{ (a) \includegraphics[width=0.43\textwidth]{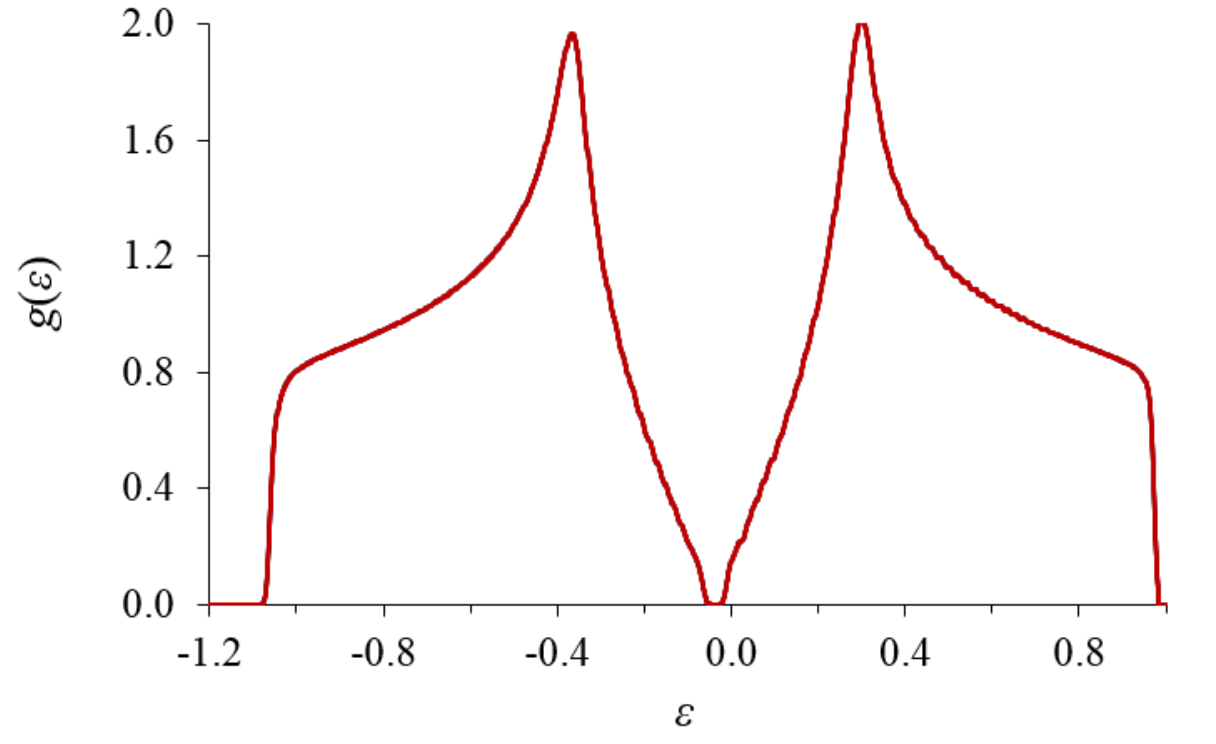} 		
								 \includegraphics[width=0.43\textwidth]{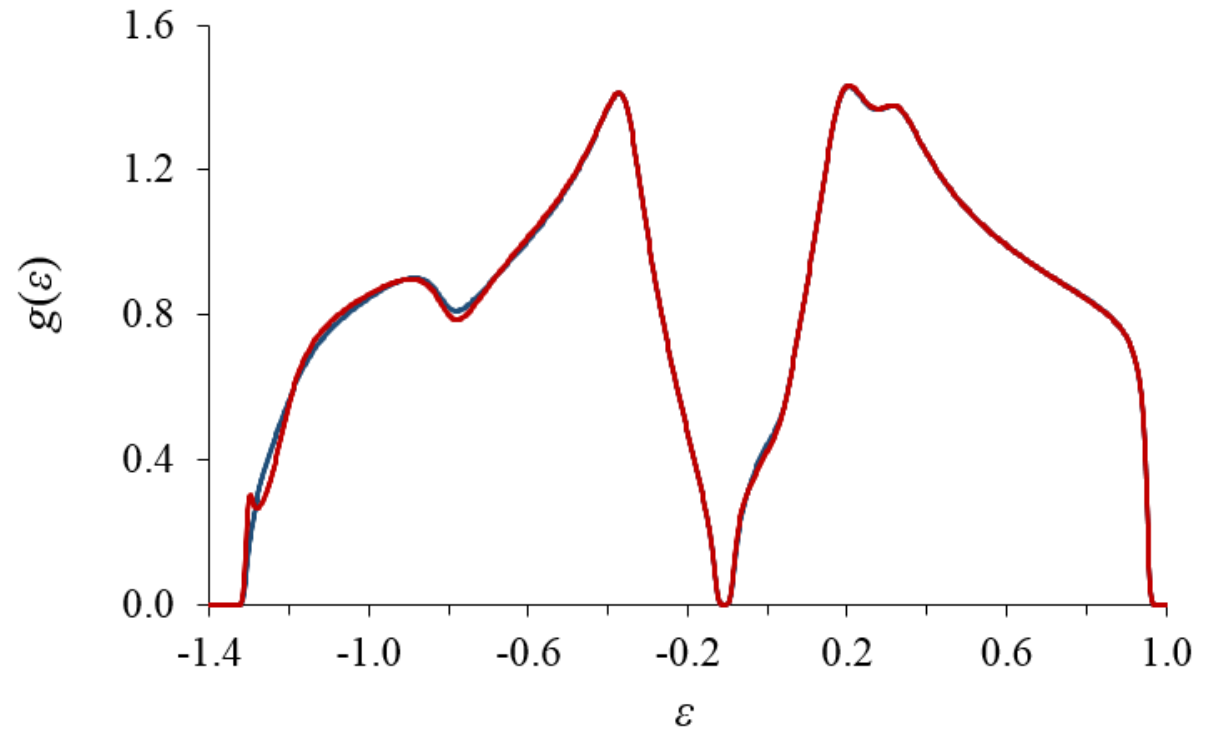}}		\smallskip
\centerline{ (b) \includegraphics[width=0.43\textwidth]{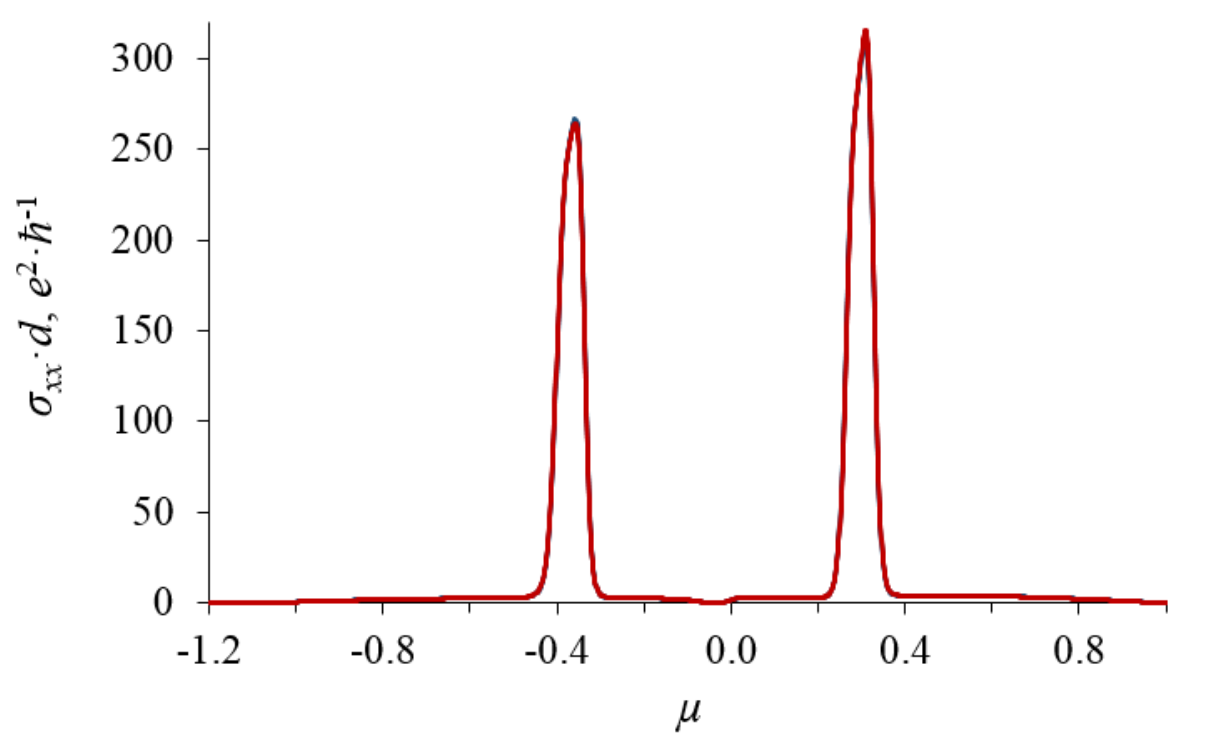} 		
								 \includegraphics[width=0.43\textwidth]{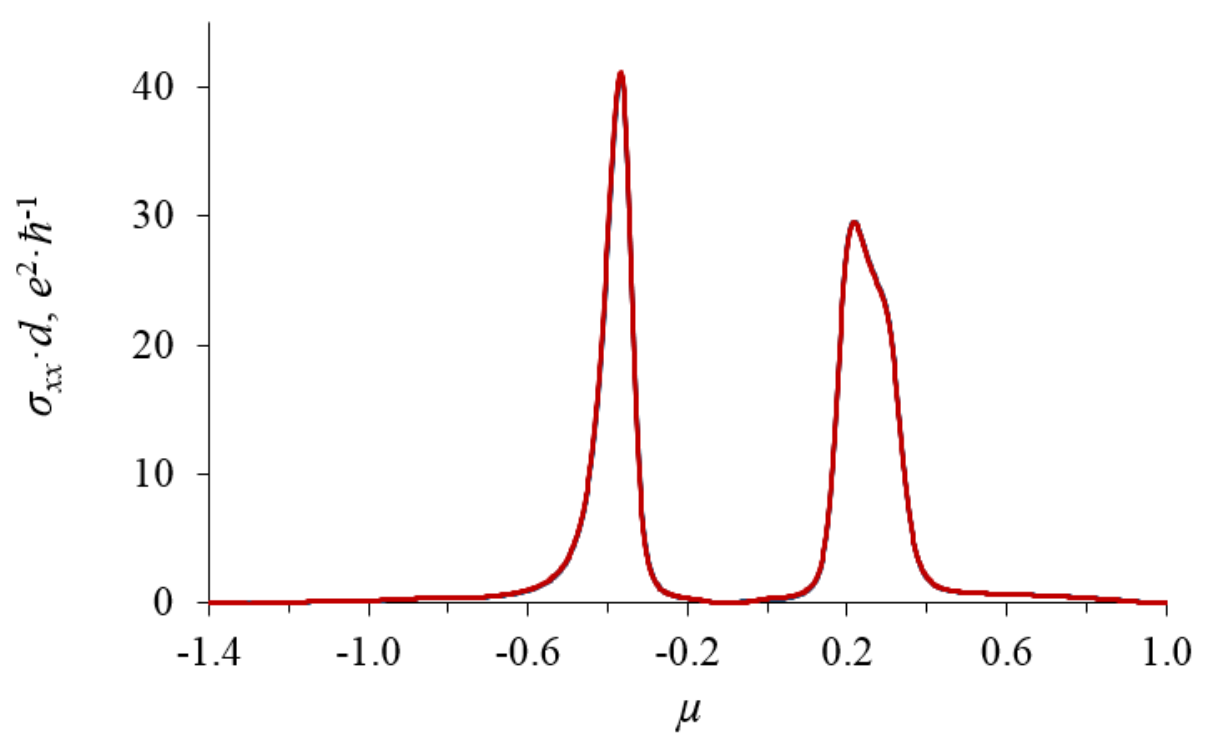}}		\smallskip
\centerline{ (c) \includegraphics[width=0.43\textwidth]{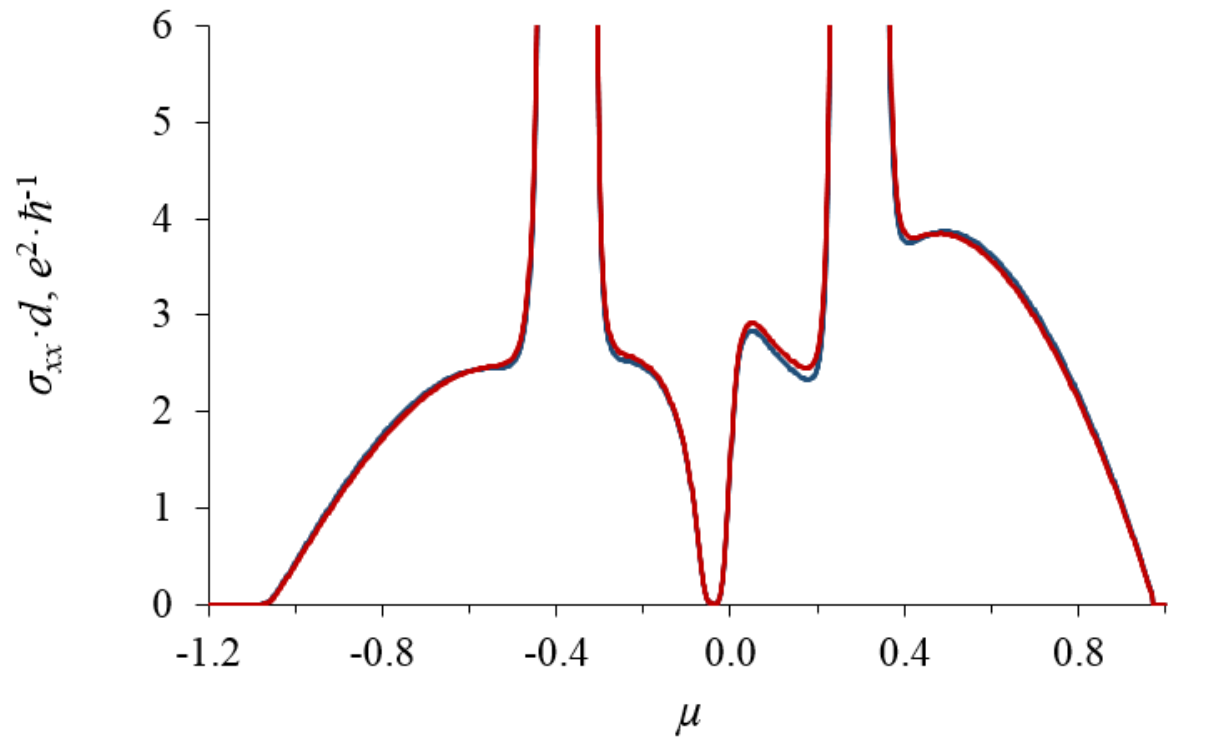} 		
								 \includegraphics[width=0.43\textwidth]{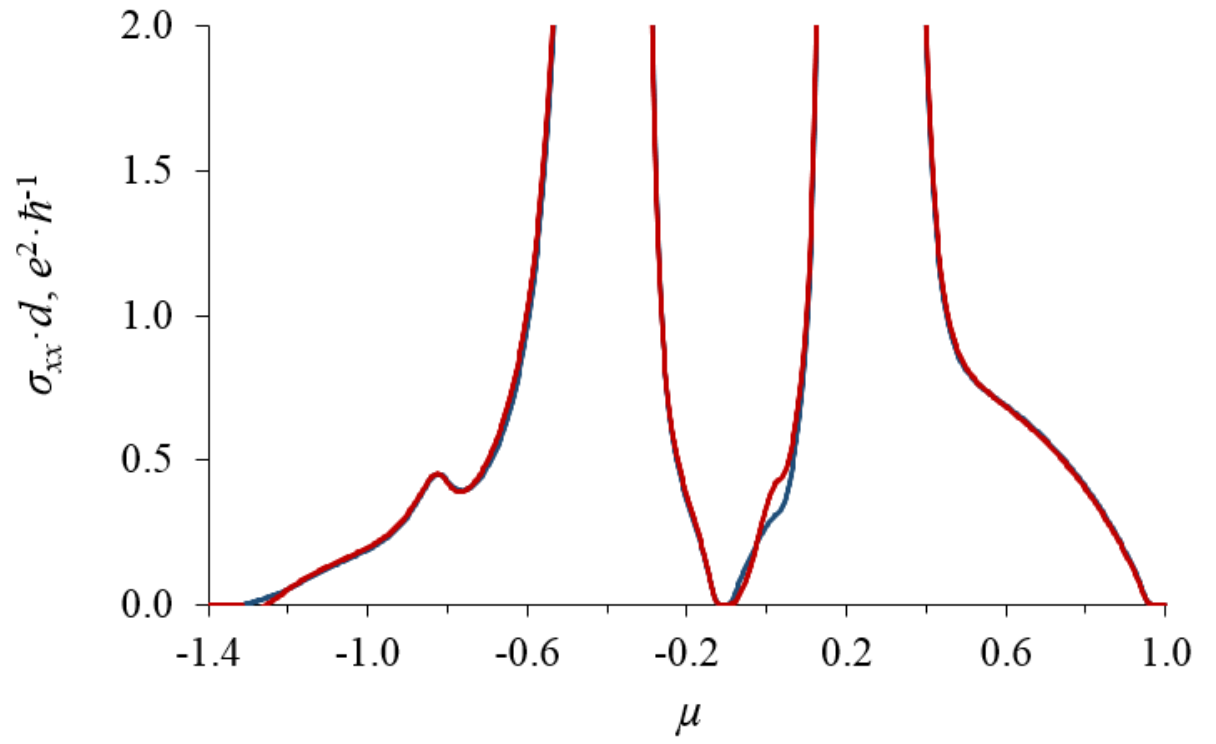}}		\smallskip
\caption{\label{fig2} (Colour online)
			(a) density of states $g(\varepsilon )$ as a function of energy $\varepsilon $,
			(b) conductivity $\sigma _{xx} (\mu) \cdot d$ as a function of Fermi level, 
					$d$ is the thickness of graphene layer; 
			(c) plot (b) enlarged on the vertical axis shown in the region of values close to the origin. 
			The scattering parameter $\delta/w=-0.2$ (lhs) and $\delta/w=-0.6$ (rhs).	
			The scattering parameter $\delta/w=-0.2$ (lhs) and $\delta/w=-0.6$ (rhs). 
			The concentration of substitutional impurity $y=0.2$, the order parameter $\eta = 0.3$. 
			Blue line presents calculations in the approximation of the coherent potential. 
			Red line, calculations taking into account the electron scattering processes 
				on the pairs of atoms of the first coordination sphere.
				} %
\end{figure}

Figure~\ref{fig2} shows a plot of the density of states $g(\varepsilon )$ vs. energy $\varepsilon $ [figure~\ref{fig2}~(a)] 
and a plot of conductivity $\sigma _{xx} (\mu )\cdot d$ vs. Fermi level $\mu $, 
$d$ is the thickness of the graphene layer [figure~\ref{fig2}~(b), figure~\ref{fig2}~(c)]. 
The energy and the Fermi level are given in the half-width energy band. 
$g(\varepsilon )$ and $\sigma _{xx} (\mu )$ calculations are executed in accordance with \eqref{Eqq8}, \eqref{Eqq9}. 
The concentration of substitutional impurity $y=0.2$, the order parameter $\eta =0.3$,
the parameter of pair interatomic correlations $\varepsilon^\text{BB}=0$, 
the scattering potential $\delta /w=-0.2$ (figure~\ref{fig2}, lhs) and $\delta /w=-0.6$ (figure~\ref{fig2}, rhs). 
$\sigma_{xx} (\mu )\cdot d$ is given in units $e^{2} \cdot \hbar ^{-1} $. 
An electrical resistance of the graphene layer
	\begin{equation} \label{Eqq33} 
	R=\frac{1}{\sigma _{xx} d} \frac{l}{L} \,,      
	\end{equation} 
where $l$ is the length of the graphene layer along the $x$-axis, $L$ is the width of the layer. The $x$-axis is directed from the carbon atom to its nearest neighbour.

With the ordering of the substitutional impurity atoms,  there is a gap in the density of states (figure~\ref{fig2}), 
the order parameter $\eta=0.3$. In the region of gap the density of states $g=0$. 
The electrical conductivity of graphene $\sigma _{xx} (\mu )$ for the Fermi level lying in the gap region is zero. 
For the Fermi level outside the gap region, the electrical conductivity of graphene differs from zero 
and increases with increasing the density of states on Fermi level.
In contrast to the case of weak scattering $|\delta /w|\ll 1$ described above, 
for which the gap width increases linearly with increasing scattering potential $|\delta /w|$, 
the dependence of the energy gap width on the case of strong scattering is more complex. 
The width of the slit decreases with increasing absolute value of the scattering potential (figure~\ref{fig2}).
The dependence of electrical conductivity on the scattering potential 
and the order parameter is also more complex than in the case of weak scattering, 
for which the electrical conductivity is described by the above formula \eqref{Eqq29}.

\begin{figure}[!t] 
\centerline{\footnotesize $\delta/w=-0.2$ \hspace{0.34\textwidth} $\delta/w=-0.6$}	\bigskip
\centerline{ (a) \includegraphics[width=0.43\textwidth]{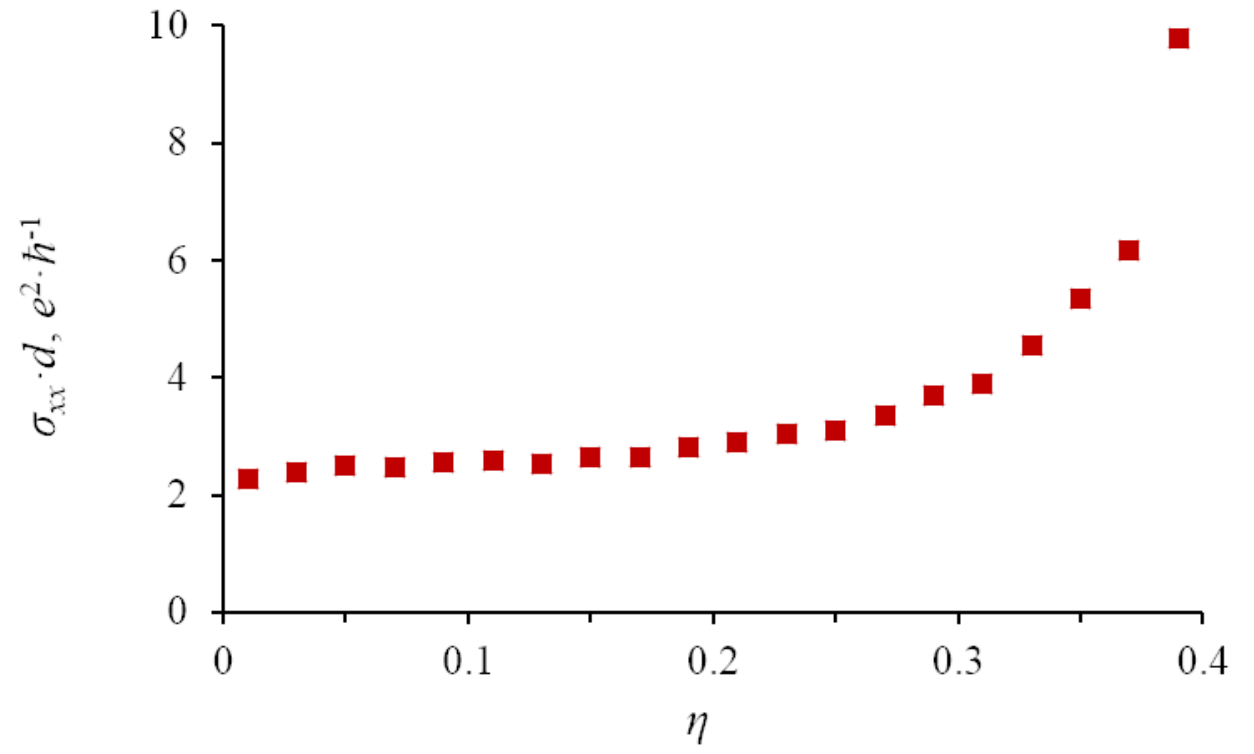} 		
								 \includegraphics[width=0.43\textwidth]{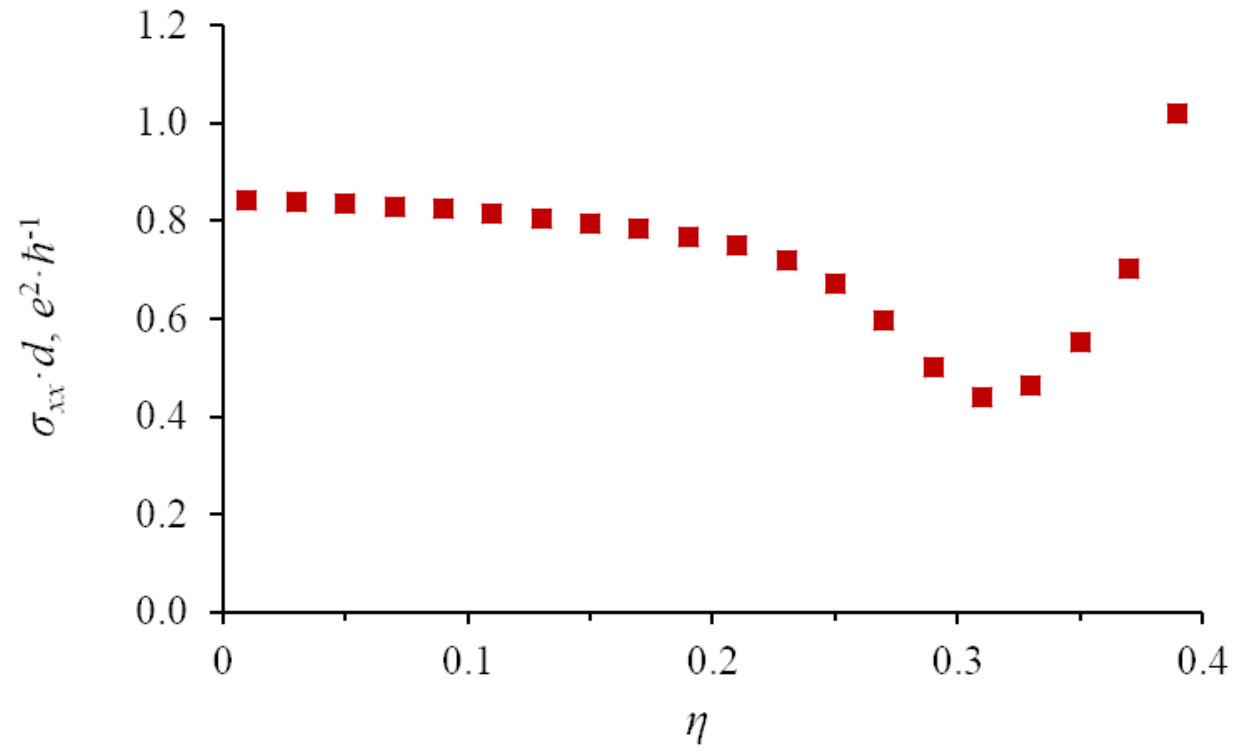}}	\bigskip
\centerline{ (b) \includegraphics[width=0.43\textwidth]{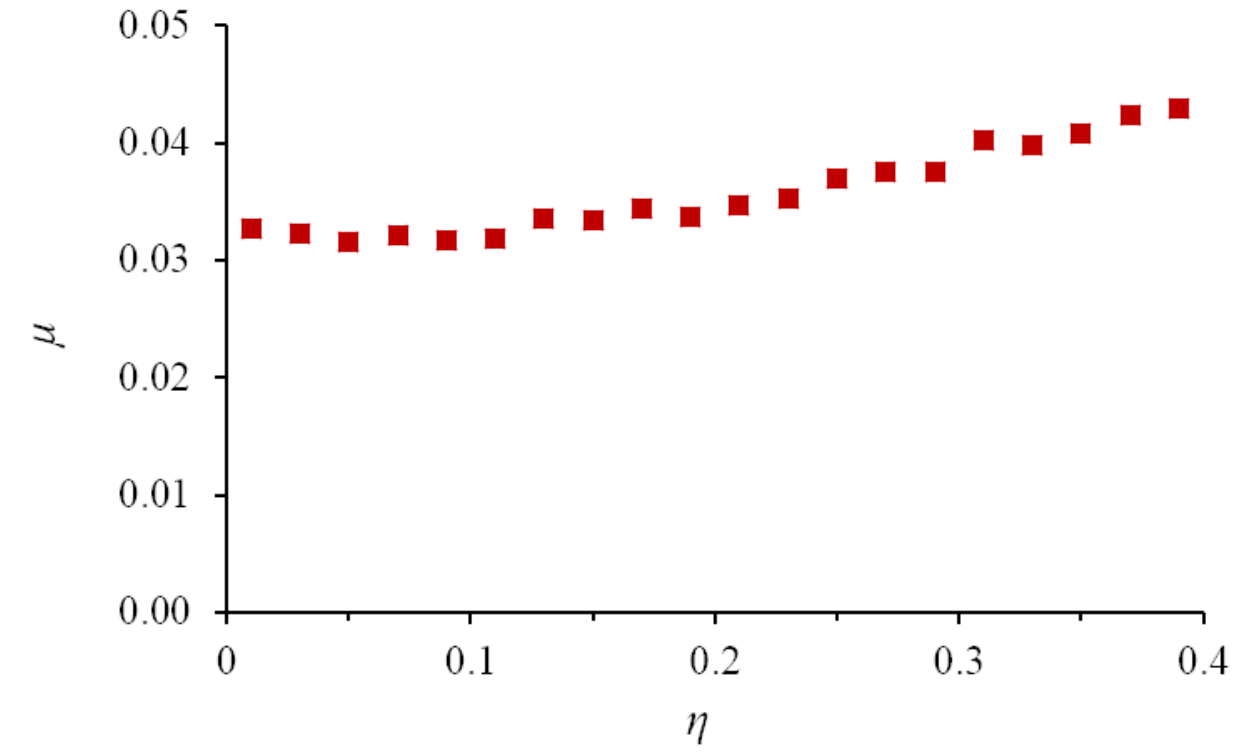} 		
								 \includegraphics[width=0.43\textwidth]{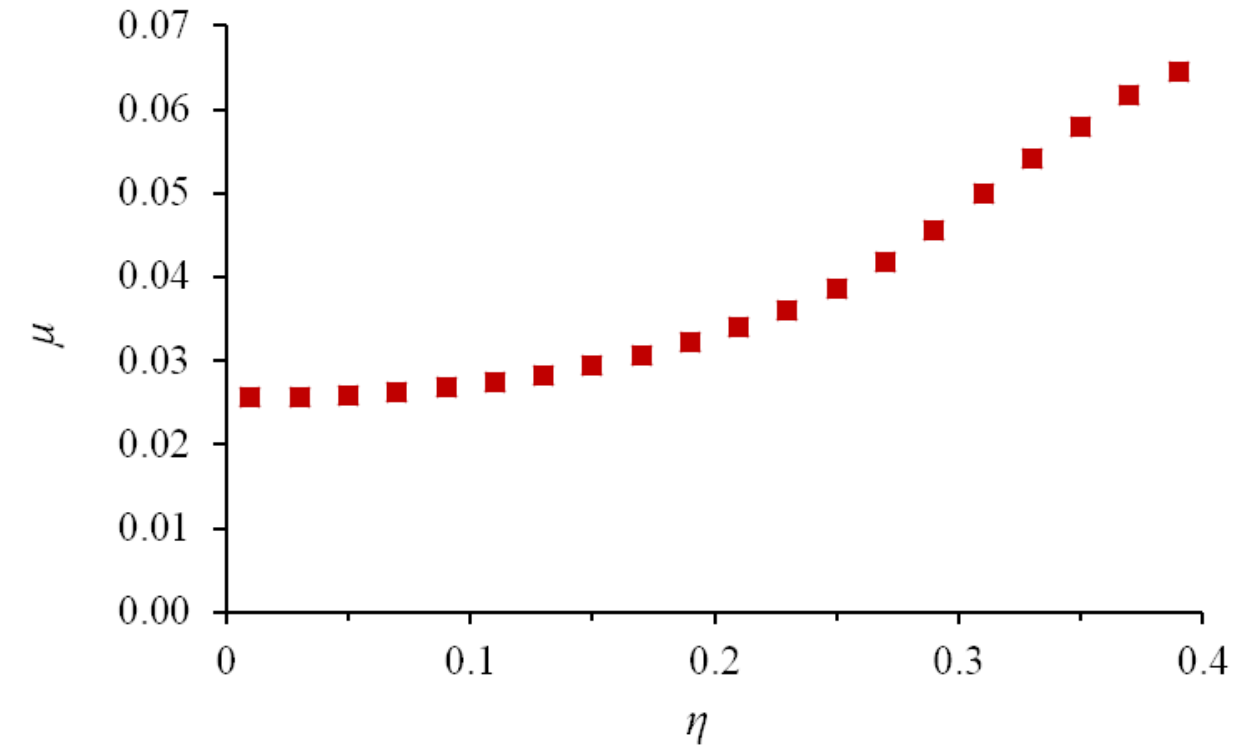}}	\bigskip
\centerline{ (c) \includegraphics[width=0.43\textwidth]{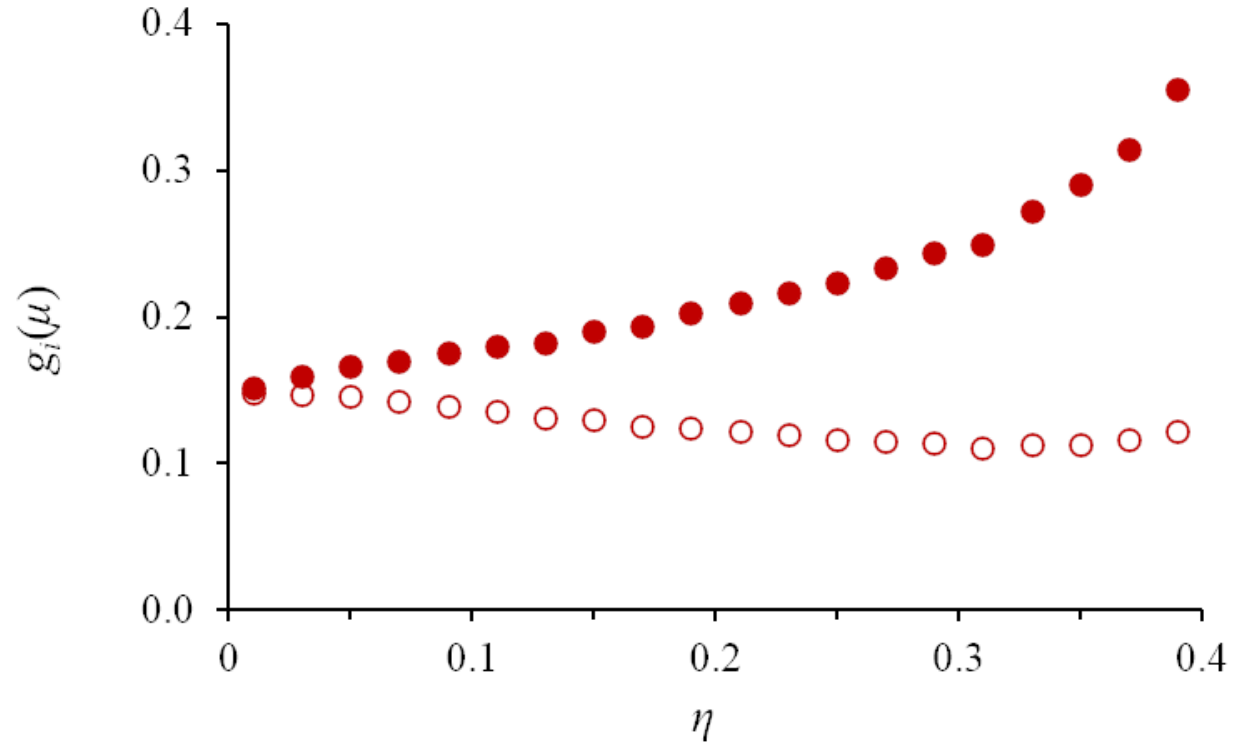} 		
								 \includegraphics[width=0.43\textwidth]{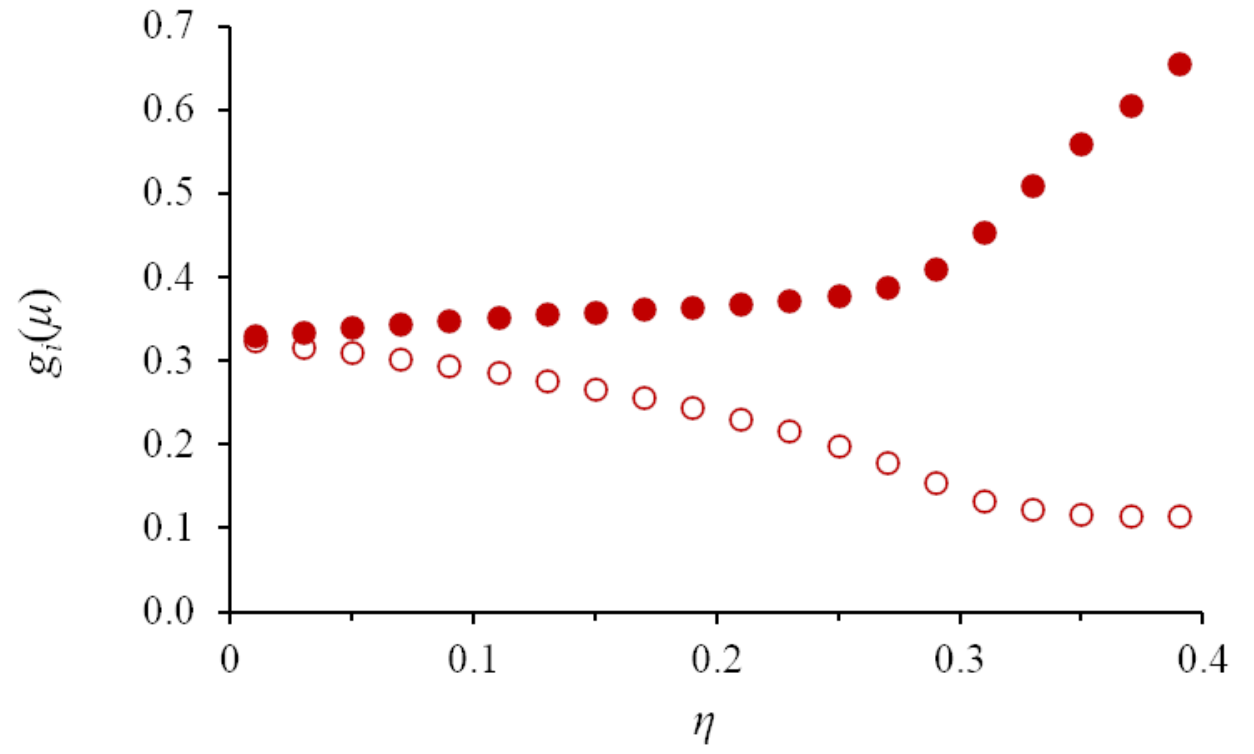}}	\bigskip
\centerline{ (d) \includegraphics[width=0.43\textwidth]{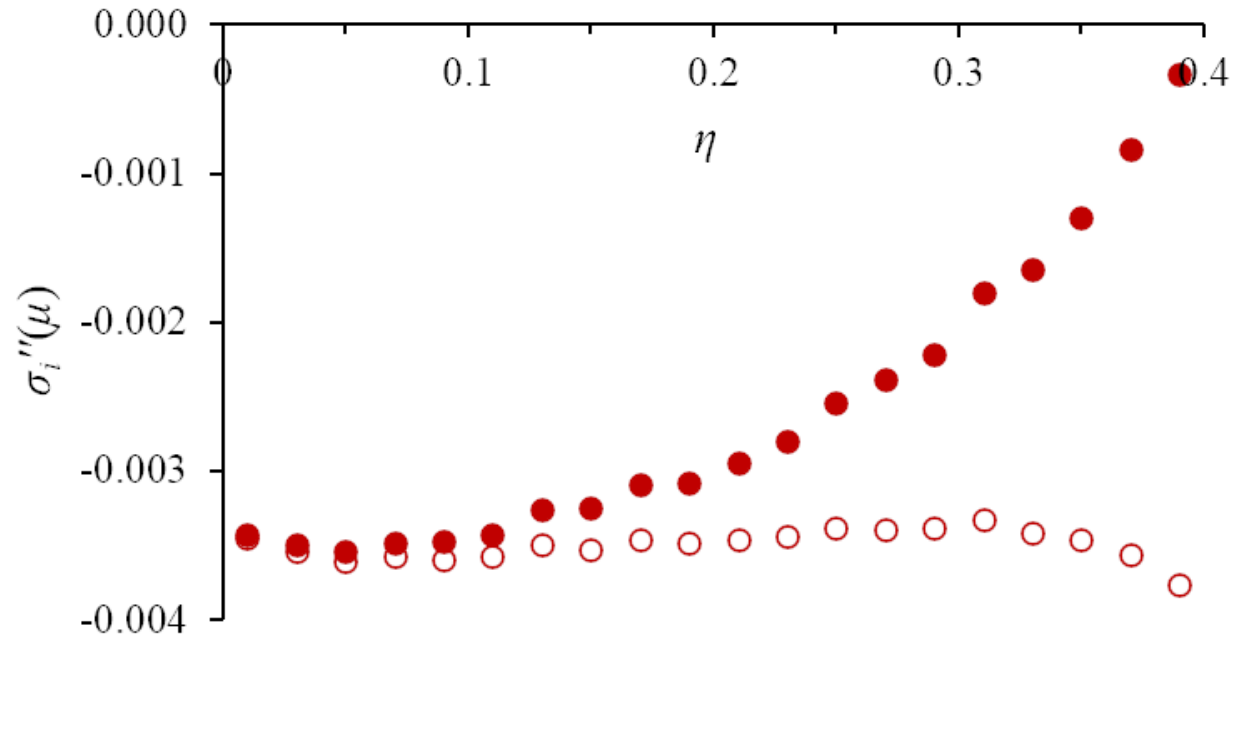} 		
								 \includegraphics[width=0.43\textwidth]{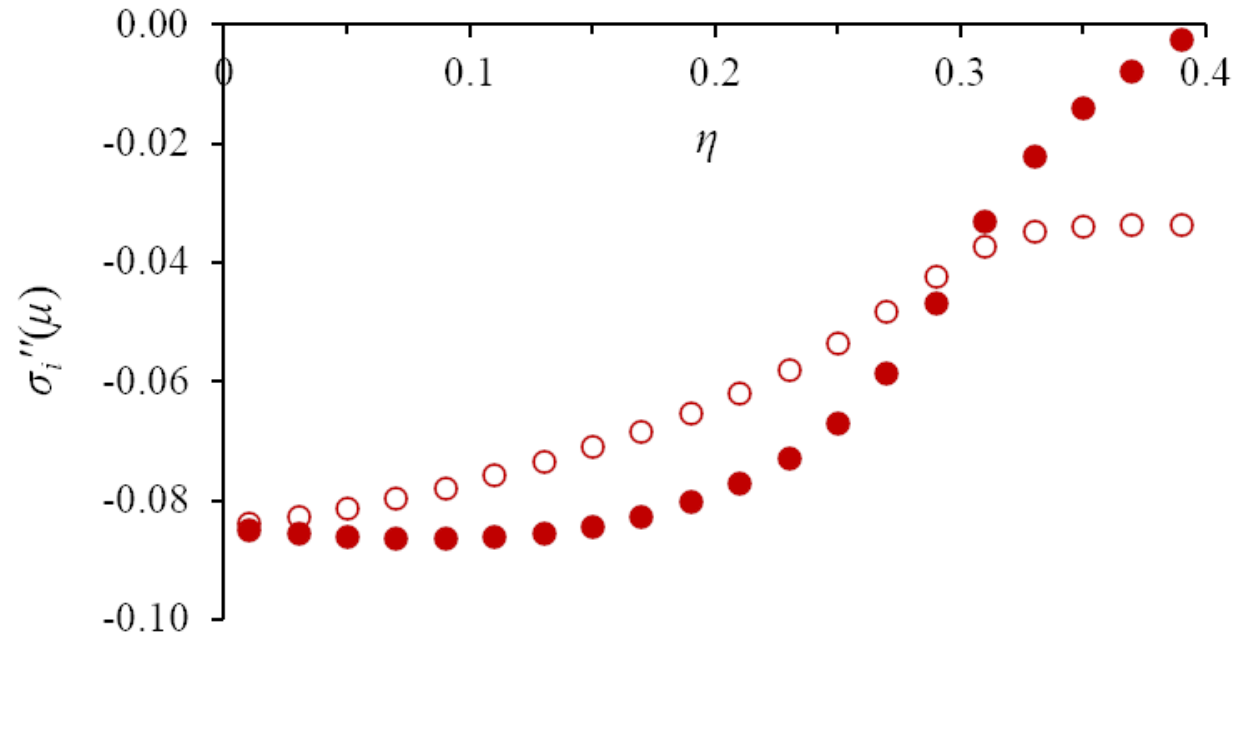}}	\smallskip
\caption{\label{fig3} (Colour online)
			(a) Conductivity $\sigma _{xx} (\mu )\cdot d$, 
			(b) Fermi level $\mu $, 
			(c) partial density of states $g_{i} (\mu )$ 
					(open circles --- $g_{1} (\mu )$, filled circles --- $g_{2} (\mu )$) 
	and (d) imaginary part of the partial coherent potential $\sigma ''_{i} (\mu )$ 
					(open circles --- $\sigma ''_{1} (\mu )$, filled circles --- $\sigma ''_{2} (\mu )$) 
			as functions of order parameter $\eta $. 
					The sublattice $i=1$ contains impurity atoms and the sublattice $i=2$ 
					contains only carbon atoms in the case of complete ordering. 
					The concentration of substitutional impurity $y=0.2$, 
					the scattering parameter $\delta/w=-0.2$ (on the left-hand side)
					and $\delta/w=-0.6$ (on the right-hand side).} 
 \end{figure}

In order to find out the nature of the dependence of the electrical conductivity 
on the values of the scattering potential $\delta$ 
and the order parameter $\eta$ on figure~\ref{fig3} are  given the electrical conductivity of graphene $\sigma_{xx}(\mu)$ 
vs. the ordering parameter of impurity atoms $\eta$ 
for different magnitude $\delta$ of the scattering potential.
The number of electrons per atom whose energy values are 
in the region of the energy band is equal $\left\langle Z \right\rangle =1.01$.
At this value $\left\langle Z \right\rangle$ the Fermi level $\mu$ lies to the right of the energy gap.
Figure~\ref{fig3}~(b) shows the Fermi level $\mu (\eta )$ depending 
on the ordering parameter of the impurity $\eta$. 
The value of the Fermi level $\mu (\eta )$ is calculated by the formula \eqref{Eqq13}. 
Figure~\ref{fig3}~(c) show the dependence of the partial density of states $g_{i} (\mu )$ 
at the Fermi level on the impurity ordering parameter $\eta$, $i$ is a sublattice number. 
Figure~\ref{fig3}~(d) show the dependence of the imaginary part 
of the coherent potential $\sigma ''_{i} (\mu )$ 
at the Fermi level on the ordering parameter $\eta$.

As can be seen from figure~\ref{fig3}, the electrical conductivity of graphene increases with increasing the ordering of the impurity  $\eta $, which is caused mainly by the increase in the density of states at the Fermi level. To find out the nature of the change in the electrical conductivity of graphene $\sigma_{xx}(\mu)$ with the change of the impurity ordering parameter  $\eta$, we turn to the limiting case of weak scattering $|\delta /w| \ll 1$.

In the coherent potential approximation for the case of weak scattering $|\delta /w|\ll 1$ we get from \eqref{Eqq9} 
\begin{equation} \label{Eqq34} 
\sigma _{\alpha \alpha } =\frac{e^{2} \hbar }{3\Omega _{0} } \sum _{i}\frac{g_{i} (\mu ){\rm \; |}\upsilon _{\alpha 12} (\mu )|^{2} }{|\sigma _{i} '' (\mu )|}  .      
\end{equation} 
For three-dimensional crystals with a simple lattice in the approximation
of the effective mass by substituting in \eqref{Eqq34} the expressions for
$g(\mu )$ and $\upsilon _{\alpha } (\mu )$, the formula takes on a
well-known form
\begin{equation} \label{Eqq35} 
\sigma _{\alpha \alpha } =e^{2} n\tau (\mu )/m^{*} ,       
\end{equation} 
$n$ is the number of electrons per unit volume whose energy is less than the Fermi level, $m^{*} $ is the electron effective mass. $\tau (\mu )$ is the relaxation time of the electronic states, which is determined by the ratio 
\begin{equation} \label{Eqq36} 
|\sigma'' (\mu )|\tau (\mu )=\hbar .       
\end{equation} 

In the case of weak scattering $|\delta /w|\ll 1$, the numerical calculation $\sigma _{xx} (\mu )$ is qualitatively in good agreement with the formulae  \eqref{Eqq30}, \eqref{Eqq34}, \eqref{Eqq29}. 
When the order parameter $\eta$ is directed to its maximum value, the electrical conductivity $\sigma _{xx} (\mu )$ tends to infinity.
As can be seen from formula \eqref{Eqq34}, this is due to an increase in the state density at the Fermi level $g_{2}(\mu)$ and an increase in the relaxation time as the order parameter $\eta$ increases (with $\eta \to \eta _{\max } = 2 y $ the imaginary part of the coherent potential $\sigma ''_{2} (\mu )\to 0$).

We note that formulae \eqref{Eqq8}, \eqref{Eqq9} for the density of states and electrical conductance of graphene cannot be used, if the Fermi level falls in the interval \eqref{Eqq32} of energies at the gap edges. 
The density of states values that are calculated taking into account the scattering processes on the pairs of atoms, which are located within the three coordination spheres and within the ten coordination spheres, coincide with the results of calculations which take into account the scattering on the pairs within the first coordination sphere. The region of impurity electronic states of graphene in the specified model is limited by the space of the first coordination sphere.

In order to describe the energy spectrum and the electrical conductivity of graphene with impurities 
we used the method of the theory of disordered systems. 
This method is based on the cluster decomposition for the one-particle Green function (state density) 
and the two-particle Green function (electrical conductivity).
For a zero one-node approximation, a coherent potential approximation is chosen 
which describes the state of the electron in some efficient ordered environment.
Herein below we present the corrections to the approximation of the coherent potential, 
which are due to the contribution of electron scattering processes on clusters of two, three, etc. atoms.
In \cite{Los14, Ducastelle19}, it has been shown that the contribution of scattering processes 
on a cluster decreases with an increase in the number of atoms in the cluster by some small parameter.
This parameter is small in the wide range of variation of the system characteristics 
except for narrow energy intervals at the edges of the spectrum and the edges of the gap \eqref{Eqq32} 
that occurs when the impurity is ordered. 
As shown in \cite{Los14, Ducastelle19} these areas \eqref{Eqq32} are the regions of localization of electronic states.
In the Van Hove regions of the energy spectrum of the pure crystal, the peaks on the state density widens due to the splitting of energy levels while reducing the symmetry of the crystal with the introduction of disordered impurity atoms. 
However, the Van Hove regions do not always coincide with the regions of localization of electronic states 
as in the case for binary alloys with BCC lattice \cite{Los14} or graphene with a substitutional impurity (figure~\ref{fig2}).

\section{Conclusions}

In the Lifshitz tight-binding one-electron model, the influence of substitutional impurity atoms on the energy spectrum and electrical conductance of graphene is studied. It is established that the ordering of substitutional impurity atoms on the nodes of the crystal lattice causes the appearance of the gap in the energy spectrum of graphene $\eta |\delta |$ in width centered at the point $y\delta $, where $\eta $ is the parameter of ordering, $\delta $ is the difference of the scattering potentials of impurity atoms and carbon atoms, $y$ is the impurity concentration. 

It is shown that if the ordering parameter $\eta $ is close to $\eta _{\max } =2y$,  $y<1/2$ then the density of electron states has peaks on the edges of the energy gap. Those peaks correspond to impurity levels. 

If the Fermi level falls in the region of the gap, then the electrical conductance $\sigma_{\alpha\alpha} \to 0$ at the ordering of graphene, i.e., the metal-dielectric transition arises. 

If the Fermi level is located outside the gap, then the electrical conductance increases with the parameter of order $\eta$. At the complete ordering of substitutional impurity $\eta \to \eta _{\max } = 2y$, the electrical conductivity $\sigma _{\alpha \alpha } \to \infty $, imaginary part of the coherent potential $\sigma ''_{2} (\mu )\to 0$ (sublattice $i = 2$ contains only carbon atoms in the case of complete ordering) and the relaxation time $\tau \to \infty $. 

The analytical expressions of the density of electron states and the electrical conductivity of graphene obtained in the case of weak scattering $\left|\delta /w\right|\ll 1$ are compared with the results of numerical calculations for a different scattering potential $\left|\delta /w\right|$, $w$ is the half-width of the energy band. The numerical calculation shows that the area of influence of impurity electronic states of graphene in the specified model is spatially limited by the first coordination sphere. 

\newpage

\section*{Acknowledgements}
S.K. acknowledges a support by the National Academy of Sciences of Ukraine by the grant
``The structure and  dynamics of statistical and quantum-field systems''
(project RK No. 0117U000240).

\newpage

\ukrainianpart

\title{Енергетичний спектр та електропровідність графену з домішкою заміщення}

\author{С.П. Репецький\refaddr{label1}, І.Г. Вишивана\refaddr{label1},  
С.П. Кручинін\refaddr{label2}, Р.М. Мельник\refaddr{label3}, А.П. Поліщук\refaddr{label4}}
\addresses{
\addr{label1} Київський національний університет імені Тараса Шевченка, Інститут високих технологій, \\просп. Академіка Глушкова, 4-г, 03022 Київ, Україна
\addr{label2} Інститут теоретичної фізики ім. М.М. Боголюбова НАН України, \\вул. Метрологічна, 14-б,	03680 Київ, Україна 
\addr{label3} Національний університет ``Києво-Могилянська академія'', вул. Г. Сковороди, 2, 04070 Київ, Україна 
\addr{label4} Національний авіаційний університет, Аерокосмічний факультет, \\просп. Космонавта Комарова, 1, 03058 Київ, Україна}
\makeukrtitle

\begin{abstract}
\tolerance=3000%
На одноелектронній моделі сильного зв'язку Ліфшиця досліджено вплив атомів домішки  заміщення  на енергетичний спектр  та  електричну провідність графену. 
Встановлено, що впорядкування атомів домішки  приводить до виникнення щілини в енергетичному спектрі електронів, ширина якої залежить від параметра порядку та величини потенціалу розсіяння. 
Показано, якщо параметр порядку близький до свого максимального значення, на кривій енергетичної залежності густини електронних станів  на краях енергетичної щілини виникають піки, пов'язані з локалізованими  домішковими станами. 
При електронній концентрації, за якої рівень Фермі попадає в область щілини,  електропровідність рівна нулю, відбувається перехід метал-діелектрик. 
Якщо рівень Фермі попадає в область енергетичної зони, час релаксації електронів і електропровідність прямують до нескінченності при прямуванні параметра порядку до свого максимального значення. 
Аналітичні розрахунки густини електронних станів та електропровідності графену, виконані в граничному випадку слабкого розсіяння, порівнюються з результатами числових розрахунків для різних потенціалів розсіяння.
\keywords графен, енергетична щілина, густина станів, параметр порядку, функція Гріна, перехід метал-діелектрик

\end{abstract}


\begin{thebibliography}{10}
\bibitem{Sun1} Sun J., Marsman M., Csonka G.I., Ruzsinszky A., Hao P., Kim Y.-S., Kresse G., Perdew J.P., 
			Phys. Rev. B, 2011, \textbf{84}, 035117, \doi{10.1103/PhysRevB.84.035117}. 
\bibitem{Yelgel2}  Yelgel C., Srivastava G.P., Appl. Surf. Sci., 
			2012, \textbf{258}, 8338--8342, \doi{10.1016/j.apsusc.2012.03.167}. 
\bibitem{Denis3}  Denis P.A., Chem. Phys. Lett., 2010, \textbf{492}, 251--257, \doi{10.1016/j.cplett.2010.04.038}. 
\bibitem{Xiaohui4} Repetsky S.P., Tretiak O.V., Vyshivanaya I.G., Ukr. J. Phys., 2015, \textbf{60}, No. 2, 170--174,\\ \doi{10.15407/ujpe60.02.0170}.
\bibitem{Skrypnyk5}  Skrypnyk Yu.V., Loktev V.M., Phys. Rev. B, 2006, \textbf{73}, 241402(R),  \doi{10.1103/PhysRevB.73.241402}.
\bibitem{Skrypnyk6}  Skrypnyk Yu.V., Loktev V.M., Phys. Rev. B, 2007, \textbf{75}, 245401,  \doi{10.1103/PhysRevB.75.245401}.
\bibitem{Pershoguba7}  Pershoguba S.S., Skrypnyk Yu.V., Loktev V.M., Phys. Rev. B, 2009, \textbf{80}, 214201, \\
			\doi{10.1103/PhysRevB.80.214201}. 
\bibitem{SkrypnykYu}  Skrypnyk Y.V., Loktev V.M., Low Temp. Phys., 2018,  \textbf{44}, 
			 1112, \doi{10.1063/1.5060964}.
\bibitem{yet1}  Eremenko V.V., Sirenko V.A., Gospodarev I.A., Syrkin E.S., Feodosyev S.B., 
			Bondar I.S., Minakova K.A., Feher A., J. Phys.: Conf. Ser., 2018, \textbf{969}, 012021, \doi{10.1088/1742-6596/969/1/012021}.
\bibitem{yet2}  Syrkin E.S., Sirenko V.A., Feodosyev S.B., Gospodarev I.A., Minakova K.A., In:
			Handbook of Graphene, Vol.~2, Stauber~T.~(Ed.), U.S. Scrivener Publishing LLC, 
			Beverly, 2019, 315--387. 
\bibitem{Radchenko8}  Radchenko T.M., Shylau A.A., Zozoulenko I.V., 
			Phys. Rev. B, 2012, \textbf{86}, 035418, \\ \doi{10.1103/PhysRevB.86.035418}. 
\bibitem{Radchenko9}  Radchenko T.M., Tatarenko V.A., Sagalianov I.Yu., Prylutskyy Yu.I., 
			Szroeder P., Biniak S., Carbon, 2016, \textbf{101}, 37--48, \doi{10.1016/j.carbon.2016.01.067}.
\bibitem{Radchenko10}  Radchenko T.M., Tatarenko V.A., Sagalianov I.Yu., Prylutskyy Yu.I.,
			Phys. Lett. A, 2014, \textbf{378}, 2270--2274, \doi{10.1016/j.physleta.2014.05.022}. 
\bibitem{Radchenko11}  Radchenko T.M., Shylau A.A., Zozoulenko I.V., Ferreira A., 
			Phys. Rev. B, 2013, \textbf{87}, 195448, \\ \doi{10.1103/PhysRevB.87.195448}.
\bibitem{Radchenko12}  Radchenko T.M., Shylau A.A., Zozoulenko I.V., 
			Solid State Commun., 2014, \textbf{195}, 88--94,\\ \doi{10.1016/j.ssc.2014.07.012}.
\bibitem{Radchenko13}  Radchenko T.M., Tatarenko V.A., Sagalianov I.Yu., Prylutskyy Yu.I., 
			Preprint \arxiv{1406.0783}, 2014.
\bibitem{Los14}  Los' V.F., Repetsky S.P., J. Phys.: Condens. Matter, 
			1994, \textbf{6}, No.~9, 1707--1730, \doi{10.1088/0953-8984/6/9/013}.
\bibitem{Repetsky15} Repetsky S.P., Vyshyvana I.G., Kruchinin S.P., Bellucci S.,  
			Sci. Rep., 2018, \textbf{8}, 9123,\\  \doi{10.1038/s41598-018-26925-0}.
\bibitem{Slater17}  Slater J.C., Koster G.F., 
			Phys. Rev., 1954, \textbf{94}, 1498--1524, \doi{10.1103/PhysRev.94.1498}. 
\bibitem{Repetsky20}  Repetsky S., Vyshyvana I., Nakazawa Y., Kruchinin S., Bellucci S., 
			Materials, 2019, \textbf{12}, 524,\\ \doi{10.3390/ma12030524}.
\bibitem{Repetsky21}  Repetsky S.P., Shatnii T.D., Teor. Mat. Fiz., 2002, \textbf{131}, No.~3, 456--478 (in Russian), \doi{10.4213/tmf340} [Theor. Math. Phys., 2002, \textbf{131}, No.~3, 832--851, \doi{10.1023/A:1015931708479}].
\bibitem{Ducastelle19}  Ducastelle F., J. Phys. C: Solid State Phys.,	1974, \textbf{7}, No.~10, 1795--1816, \doi{10.1088/0022-3719/7/10/007}.


\end{thebibliography}
\end{document}